\documentclass[%
reprint, 
superscriptaddress, 
amsmath, amssymb, 
aps, 
pra,
longbibliography]{revtex4-1}
\usepackage{threeparttable} 
\usepackage{booktabs} 
\usepackage{graphicx}
\usepackage{amsfonts}    
\usepackage{verbatim}   
\usepackage{color}      
\usepackage{subfigure}  
\usepackage{hyperref}   
\usepackage{booktabs}
\usepackage{threeparttable}
\usepackage{dcolumn}
\usepackage{multirow}
\usepackage{physics}
\usepackage{amsmath,bm} 
\usepackage{appendix} 

\usepackage[ruled]{algorithm2e}

\newcommand{\myExpect}[1]{\langle #1 \rangle} 

\begin{document}

\title{ Fast qubit-based frequency recovery algorithm for quantum key distribution  }

\author{Feng-Yu Lu}\email{These authors contributed equally to this work}
\author{Zheng-Kai Huang}\email{These authors contributed equally to this work}
\author{Jia-Jv Deng}
\affiliation{CAS Key Laboratory of Quantum Information, University of Science and Technology of China, Hefei, Anhui 230026, P. R. China}
\affiliation{CAS Center for Excellence in Quantum Information and Quantum Physics, University of Science and Technology of China, Hefei, Anhui 230026, P. R. China}
\affiliation{Hefei National Laboratory, University of Science and Technology of China, Hefei 230088, China}
\author{Chi Zhang}
\affiliation{CAS Key Laboratory of Quantum Information, University of Science and Technology of China, Hefei, Anhui 230026, P. R. China}
\affiliation{CAS Center for Excellence in Quantum Information and Quantum Physics, University of Science and Technology of China, Hefei, Anhui 230026, P. R. China}
\author{Shuang Wang}\email{wshuang@ustc.edu.cn}
\author{De-Yong He}\email{hedeyong@mail.ustc.edu.cn}
\author{Zhen-Qiang Yin}
\author{Wei Chen}
\author{Guang-Can Guo}
\author{Zheng-Fu Han}
\affiliation{CAS Key Laboratory of Quantum Information, University of Science and Technology of China, Hefei, Anhui 230026, P. R. China}
\affiliation{CAS Center for Excellence in Quantum Information and Quantum Physics, University of Science and Technology of China, Hefei, Anhui 230026, P. R. China}
\affiliation{Hefei National Laboratory, University of Science and Technology of China, Hefei 230088, China}
\date{\today}

\begin{abstract}
    Clock synchronization serves as a foundational subsystem in quantum key distribution (QKD). The recently proposed Qubit-based synchronization (Qubit4Sync) has opportunities in eliminating additional cost, noise, and potential side channels. It offers a promising alternative to dedicated synchronization hardware. However, the current frequency recovery process in Qubit4Sync requires high data throughput and computational speed, limiting practical use. To overcome these issues, we developed a fast frequency recovery algorithm that increases the recovery rate by orders of magnitude and remains robust under bad signal-to-noise ratio (SNR). This enables Qubit4Sync to operate effectively in mainstream gated-mode QKD systems. We further establish a theoretical model for frequency recovery, showing that our algorithm is robust against disturbances like dead time, jitter, and afterpulse. A frequency-domain SNR calculation method is also provided to guide parameter design for specific experimental conditions. This work opens the door to practical Qubit4Sync deployment in general QKD systems.
\end{abstract}

\maketitle

\section{INTRODUCTION} 

Quantum key distribution (QKD) \cite{bb84} allows two remote users, named Alice and Bob, to establish a secret key sharing with information-theoretic security over an insecure channel. Since the first QKD was performed at a 30 cm distance table-top experiment in 1992 \cite{bennett1992experimental}, researchers have successfully implemented QKD across diverse platforms and scenarios, such as satellites \cite{liao2017satellite}, drones \cite{tian2024experimental}, fiber links \cite{li2023high,lu2023experimental,hu2023proof}, free-space links \cite{schmitt2007experimental,vallone2015adaptive,avesani2021full,chen2021integrated}, and quantum networks \cite{chen2021integrated,joshi2020trusted,wengerowsky2018entanglement,fan2021measurement,fan2022robust}. These advancements have achieved breakthroughs in critical metrics such as long-distance transmission \cite{wang2022twin,liu2023experimental}, high key rates \cite{li2023high,grunenfelder2023fast}, and robust system stability \cite{agnesi2020simple,pereira2020quantum,lu2022unbalanced}. Behind these achievements lies a common basis: a stable and reliable synchronization. The reason is that the foundational security requires Alice and Bob to modulate the quantum states randomly and independently. Without a precise common time reference, misalignment between preparation and detection will occur, resulting in excessively high quantum bit error rates. Consequently, excessive key bits are consumed in error correction and privacy amplification, leading to the failure of key distribution.

Up to now, nearly all existing QKD systems synchronize their clocks with additional hardware such as global navigation satellite systems \cite{schmitt2007experimental,vallone2015adaptive,avesani2021full}, electrical cables \cite{wang2019beating,wang2022twin,liu2023experimental,lu2023experimental,hu2023proof}, or timing reference laser \cite{wang2014field,liu2021field,li2023high,grunenfelder2023fast}. However, the hardware increases resource overhead, introduces background noise, and more seriously, may bring potential loopholes \cite{qi2005time,zhao2008quantum}.
 As a countermeasure, methods for removing the hardware synchronization were explored. The first qubit-based synchronization algorithm was proposed and named Qubit4Sync \cite{agnesi2020simple,calderaro2020fast} in 2020. After that, the following works have progressively refined the Qubit4Sync to make it more flexible and robust \cite{cochran2021qubit,mantey2022frame,chen2024qubit,sun2025optimized}, and extended its applications to more protocols \cite{huang2024qubit} and systems \cite{scalcon2022cross,agnesi2024field,picciariello2025intermodal}. Usually, a Qubit4Sync algorithm consists of a frequency recovery and a time-offset recovery. In the frequency recovery stage, Bob employs Fourier analysis (typically, the fast Fourier transform, FFT) and statistical algorithms to deduce Alice's clock frequency. Since the clock information is hidden in Alice's periodic pulses, Bob can post-process the pulse arrival time, then deduce and recover his clock frequency according to the post-processing result.
 
However, the aforementioned works impose strict system requirements, significantly hindering the practical deployment, especially in low-power and miniaturized systems. The previous frequency recovery stage requires high-frequency sampling on the SPD counts. According to the Nyquist sampling theorem, Bob’s sampling rate must exceed twice the system frequency. Such high-frequency sampling imposes significant challenges on system bandwidth, data throughput, and the processing speed of FFT operations. Additionally, nearly all current Qubit4Sync experiments are realized in free-running systems \cite{lunghi2012free,korzh2014free} --- in QKD setups using gated-mode SPDs \cite{he2017sine,jiang20171}, any frequency mismatch between Alice and Bob causes a substantial proportion of qubits to fall outside the detection time windows, thereby reducing the system's signal-to-noise ratio (SNR). Nevertheless, gated-mode SPDs remain the predominant architecture in commercial QKD systems. Intuitively, the inherent low-SNR prevents people to employ Qubit4Sync in gated-mode systems.


To address the challenges and generalize the Qubit4Sync, we designed a fast frequency recovery algorithm that improves the frequency recovery rate by several orders of magnitude and performs robustly even under low SNR and high jitter, afterpulse, and dead time conditions, thus unlocking the implementation in gated-mode QKD systems, which is still the mainstream in QKD's application. Additionally, we presented a theoretical framework that comprehensively incorporates the practical issues into the theory, quantifies the impact of various practical issues, and builds a method for calculating the frequency-domain SNR, which can help guide users to design Qubit4Sync parameters for their specific QKD setups. Finally, we employed Monte Carlo simulations to evaluate the performance of our fast frequency recovery method in both 20 MHz and 1 GHz QKD systems. The results demonstrate that, even under low qubit count rate and in the presence of various impairments such as jitter, afterpulse, and dead time, our approach achieves accurate frequency calibration within a significantly shorter time. To sum up, this work addresses the critical issues in the original Qubit4Sync algorithms --- excessive data requirements and slow computation in frequency recovery --- while also filling gaps in its theoretical foundation. It demonstrates that Qubit4Sync can perform effectively even in low-SNR gated systems, thereby paving the way for its practical deployment in real-world QKD implementations.


The remainder of this paper is organized as follows. In Sec. \ref{sec II}, we introduce our fast frequency recovery algorithm and compare it with the previous methods. Then in Sec. \ref{sec III}, we simulate the QKD process by Monte Carlo and validate its high performance. Finally, in Appendix \ref{sec IV}, we present our theoretical foundation for Qubit4Sync.

\section{Qubit4Sync and fast frequency recovery algorithm}

\label{sec II}

\subsection{Frequency recovery algorithm for Qubit4Sync}

\begin{figure}[htbp]
	\includegraphics[width=8cm]{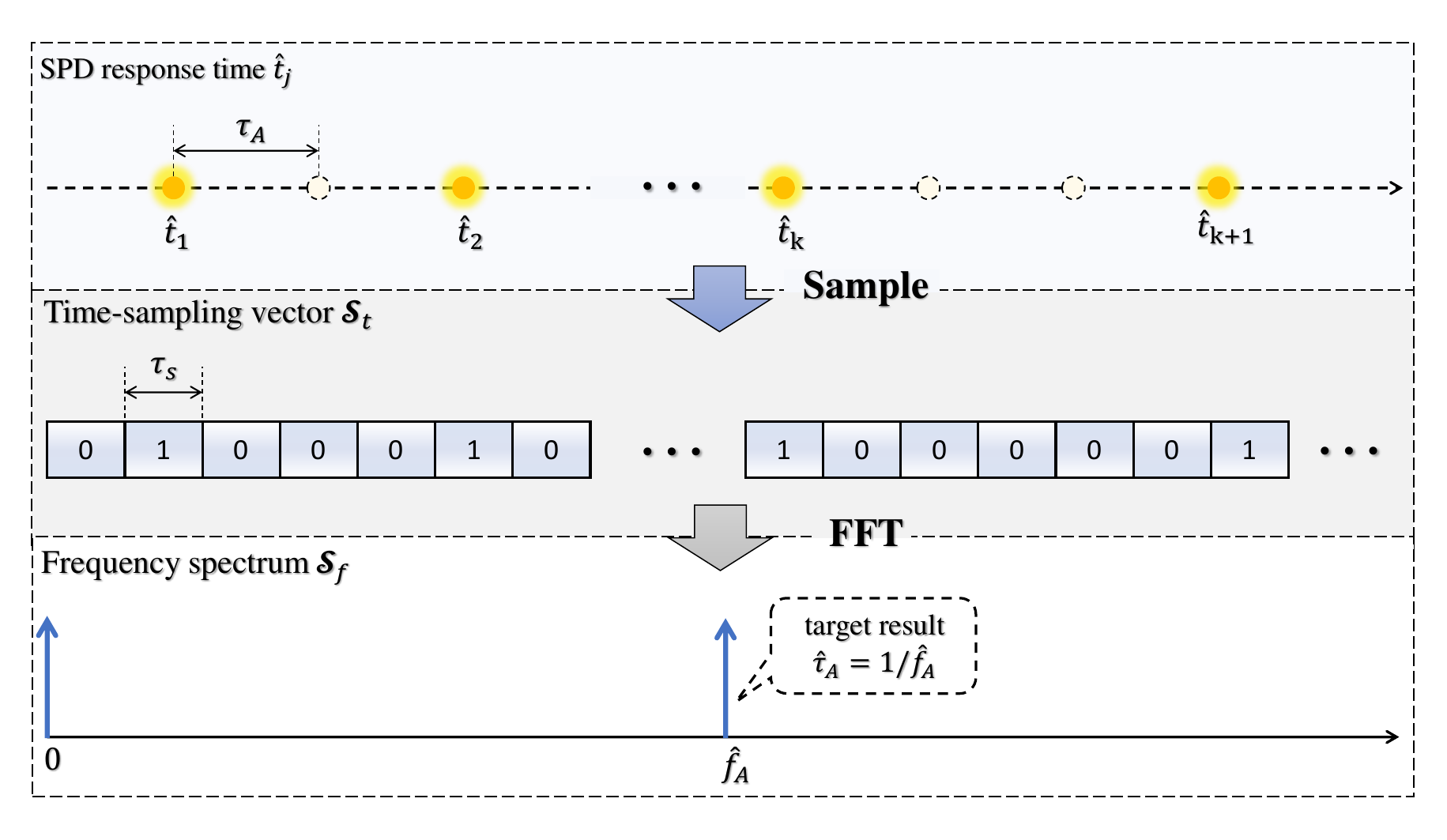}
	\caption{\label{fig: original_Qubit4Sync} {\bf Diagram of the Fourier analysis in previous Qubit4Sync.} (a) Procedure of Bob's detection: The dashed arrow indicates the timeline from Bob's perspective. Solid yellow circles represent pulse arrivals while Bob's free-running SPD detects a response. Dashed circles denote pulse arrivals without SPD-click. (b) Bob's sample for the Fourier analysis: Bob divides the timeline into multiple intervals and digitizes each interval based on the presence or absence of an SPD click. (c) The frequency spectrum obtained by the FFT. The blue arrow indicates Alice's system frequency.}
\end{figure}

First of all, we briefly introduce the original frequency recovery in Qubit4Sync, whose mission is to help Bob estimate Alice's period $\tau_A$ (equivalent to estimating frequency $f_A = 1 / \tau_A $). Bob selects one of his \textit{free-running SPD} for synchronization and estimates $\tau_A$ through its detection events. As illustrated in Fig. \ref{fig: original_Qubit4Sync}, Bob records the $l^{\text{th}}$ count time as $\hat{t}_l$ and samples the time domain with a \textit{high sample-rate} $f_s$, which must exceed twice the system frequency to satisfy the Nyquist sampling theorem. Defining $\tau_s = 1 / f_s$ as the sample period, $T_s = N\tau_s$ as the sample time, and array $\bm{x}$ as the sample result. When it satisfies $\hat{t}_l \in [(l-1)\tau_s,l\tau_s)$, Bob records the $l^{\text{th}}$ element $x(l)=1$, and the other sample results are recorded as 0. A spectrum array $\bm{X}$ can be obtained by employing the fast Fourier transform (FFT) to $\bm{x}$. The index of the first peak $k_P$ ($k_P \neq  0$) in the vector $\bm{X}$ indicates a coarse clock frequency estimation $\hat{f}_{coa} = k_p/T_s$. $\hat{\tau}_{coa} = 1 \big/ \hat{f}_{coa}$ denotes the estimated period, and $\Delta_{\tau_{coa}} = \tau_A - \hat{\tau}_{coa}$ is the target of the following fine-tuning.

Based on the coarse-tuning, statistical algorithms \cite{agnesi2020simple,calderaro2020fast,chen2024qubit,huang2024qubit} are available for higher precision. Bob estimates the round index and accumulated error as 
\begin{equation}
\begin{aligned}
\label{eq 22222}
    &k_j = \lfloor (\hat{t}_j - \hat{t}_0) /  \hat{\tau}_{coa} \rfloor, \ \ \text{and}\\
    &y_j = (\hat{t}_j - \hat{t}_0) \mod \hat{\tau}_{coa},
\end{aligned}
\end{equation}
 It enlightens us that regression algorithms may be suitable for further estimation. Taking the least squares regression (LSR) as an example, the fine-tuning is expressed as 
\begin{equation}
\label{eq LSR}
\begin{aligned}
  \hat{\tau}_{fin} = \sum_j (k_j - \bar{k}) ( y_j - \bar{y} ) \big/   \sum_j (k_j - \bar{k}),
\end{aligned}
\end{equation}
where $\bar{k} = \sum_j k_j / \max(j)$ and $\bar{y} = \sum_j y_j / \max(j) $ respectively. Finally, Bob recovers his period as $\tau_B = \hat{\tau}_{coa} + \hat{\tau}_{fin} $, which is sufficiently close to $\tau_A$.

\subsection{Fast frequency recovery algorithm}

\begin{figure}[htbp]
	\includegraphics[width=8cm]{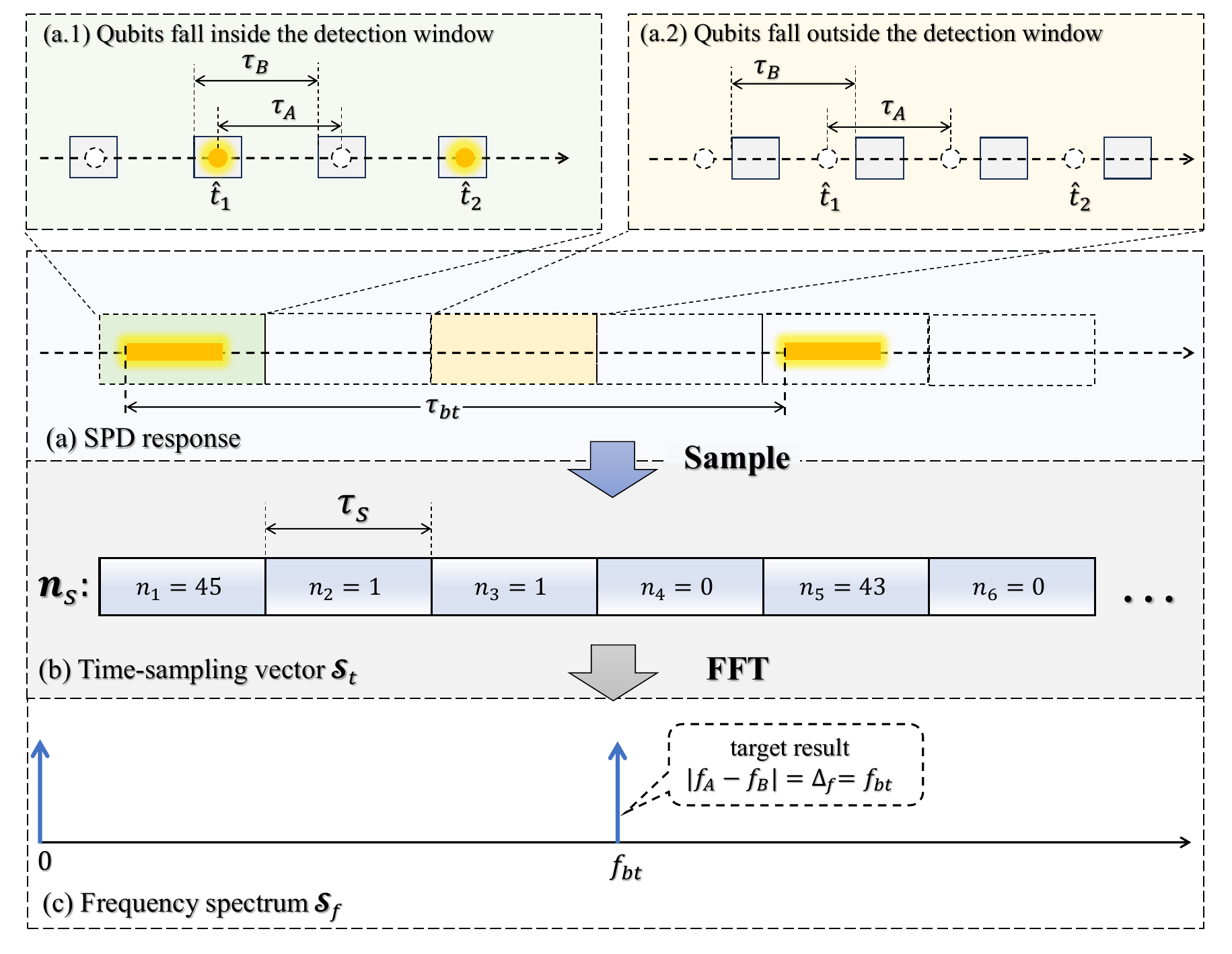}
	\caption{\label{fig: fast_qubit4sync} {\bf Diagram of the Fourier analysis in our fast frequency recovery.} (a) Frequency-mismatched gate-mode detection: The dashed arrow indicates the timeline from Bob's perspective. Yellow rectangle region represents the time interval during which qubits fall inside the detection window. Other regions without the yellow rectangle represent the time interval during which qubits fall outside the detection window. (a.1) and (a.2) represent magnified views of the corresponding regions, in which solid yellow circles and dashed circles denote SPD count with and without SPD-click respectively, the solid blue boxes represent the detection windows. (b) Bob's sample for Fourier analysis: Bob divides the timeline into many intervals and counts their count number. (c) The frequency spectrum obtained by the FFT. The blue arrow indicates the beat frequency.}
\end{figure}

A critical issue is that high-speed sampling generates a substantial volume of data, posing significant challenges to system throughput and the computational speed of Fourier analysis, which contradicts the trend toward miniaturization and integration in QKD systems. Moreover, nearly all existing implementations of Qubit4Sync have been confined to free-running detectors, which contradicts the widespread adoption of gated single-photon avalanche photodiodes (SPAD) in practical systems. The primary reason for avoiding Qubit4Sync in gated mode is that any frequency mismatch between Alice and Bob causes a large portion of qubits to fall outside the detection window, drastically reducing the SNR. Since algorithm-based calibration usually requires a high SNR, it is intuitively assumed that frequency mismatch would severely undermine frequency recovery performance in gated operation. However, we find that such a mismatch also introduces distinctive features in the frequency domain, which unexpectedly enables the success of our fast frequency recovery method.

Let us briefly introduce the key idea of the fast frequency recovery. As illustrated in Fig. \ref{fig: fast_qubit4sync}, in frequency-mismatched gated-mode detection, qubits from Alice periodically fall in and out of the detection window, leading to a cyclic variation in the count rate --- a beating phenomenon. The frequency of this variation, i.e., the beat frequency $f_{bt}$, equals the frequency difference between Alice and Bob $\Delta_f =f_A - f_B$, thereby enabling frequency recovery through its measurement. Since $f_{bt}$ is substantially lower than the QKD system frequency $f_A$ ($f_B$), low-speed sampling becomes feasible. This approach offers several key advantages: it greatly reduces the amount of data for transmission and processing, easing the burden on readout electronics; it remains robust against non-ideal detector effects—such as dead time, afterpulse, and jitter --- which typically impair high-frequency sampling; and it is applicable to both gated-mode and free-running single-photon detectors, where in the latter case the beat can be synthetically generated via post-processing. Our frequency recovery can be regarded as the combination of several modules as follows:\\


\begin{figure*}[htbp]
    \includegraphics[width=13cm]{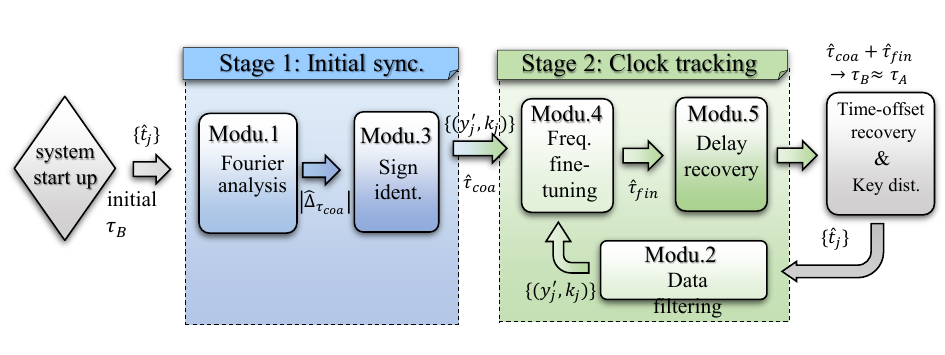}
        \caption{\label{fig: our sync algorithm} \textbf{Procedure of our frequency recovery.} When a system is initially started up. Bob should employ our fast frequecy recovery for estimating a coarse period. Then he periodically performs the frequency fine-tuning, time-offset recovery, and sends qubits to realize the clock tracking. }
    \end{figure*}

\textbf{Module 1. Fourier analysis}: In this module, Alice and Bob pre-decided their expected precision $\Delta_{f_{s}}$ according to the knowledge of their own clocks and systems. The two users maintain their QKD continuously operated under the misaligned clocks for $T_s$ duration, where the $T_s$ is actually the sample duration that should satisfy $T_s \geq 1 \big/ \Delta_{f_{s}} $. Bob records all time tags $\hat{t}_j$ for detection events that occur within the sample duration. When free-running detection is employed, Bob first filters his time tags using a digital gating sequence based on an initially estimated frequency $f_B$, thereby synthetically generating a beat frequency. In the case of gated-mode operation (here $f_B$ corresponds to the actual gate repetition rate), the beat frequency arises naturally. Then, for estimating the beat frequency, 
Bob divides the sample duration to $N_s = T_s \big/ \tau'_s$ time intervals, where $\tau'_s = 1 / f'_s$ is the time interval duration and $f'_s$ can be regarded as the sampling rate. In our method, it only requires a low-speed sampling of $f'_s \geq 2 \Delta_{f}$.
We Define $\mathbb{T}_l$ as a set that consists of all time tags $\hat{t}_j \in [(l-1)\tau_s,l\tau_s)$, Bob counts the time tag number $ x (l) = |\mathbb{T}_l|$ of each interval, and arrange these $x(l)$ into an array ${\bm x}$ ordered by $l$. By employing the FFT to the array $\bm x$, Bob obtains an array $\bm{X}$ representing the spectrum. Its first peak index $k_p$ indicates the beat frequency $  f_{bt} \approx k_p\Delta_{f_{s}} = |\hat{\Delta}_{f_{coa}}|$. The absolute period difference is estimated by $|\hat{\Delta}_{\tau_{coa}}| = \tau_B| \hat{\Delta}_{f_{coa}}| / f_B$.\\

 \textbf{Module 2. Data filtering}:
    Equation (\ref{eq 22222}) indicates an approximate linear model that 
    \begin{equation}
        \label{eq: linear model}
        y_j =  ( k_j \Delta_{\tau_{coa}}  + \varepsilon_j - \varepsilon_0 ) \mod \tau_{coa} , 
    \end{equation}
    where $\varepsilon_j$ is the random error of the $j^{\text{th}}$ round. However, we observed that the random noise $\varepsilon_j - \varepsilon_0$ would dramatically break the linear property if $k_j \Delta_{\tau_{coa}}  \approx 0$. So, compared with the previous Qubit4Sync, we additionally add a data filtering as follows for improving the algorithm robustness. First of all, Bob finds a $\tilde{j} \approx \max(j)/2$ and pre-processes all his time tags by
    \begin{equation}
     \label{eq: new dataset prepross}
    \begin{aligned}
        &t'_j = \hat{t}_j - \hat{t}_{\tilde{j}} + \tau_{coa}/2, \ \ \text{and} \\
        &y'_j = t'_j \mod \tau_{coa}, \\
        & k_j = \lfloor ( t_j - t_0 ) / \tau_{coa} \rfloor
    \end{aligned}
    \end{equation}
     which guarantees $y'_{\tilde{j}} = \tau_{coa}/2$ so that most of the $y'_j \mod \tau_{coa}$ falls in the central of the interval $(0,\tau_{coa})$. As we know $y'_j$ near $0$ and $\tau_{coa}$ are sensitive to noise, we delete all these $y'_j$ and redefine a filtered set
        \begin{equation}
     \label{eq: filtered set}
    \begin{aligned}
        \mathbb{J} = \{j \Big| |y'_j| > \epsilon_{th} \cup |y'_j - \tau_{coa}| > \epsilon_{th}  \},
    \end{aligned}
    \end{equation}
    where the elements in the set $\mathbb{J}$ are arranged in \textit{ascending order} and $\epsilon_{th}$ is a pre-decided threshold depending on the noise level.\\

\begin{figure}[htbp]
\includegraphics[width=8cm]{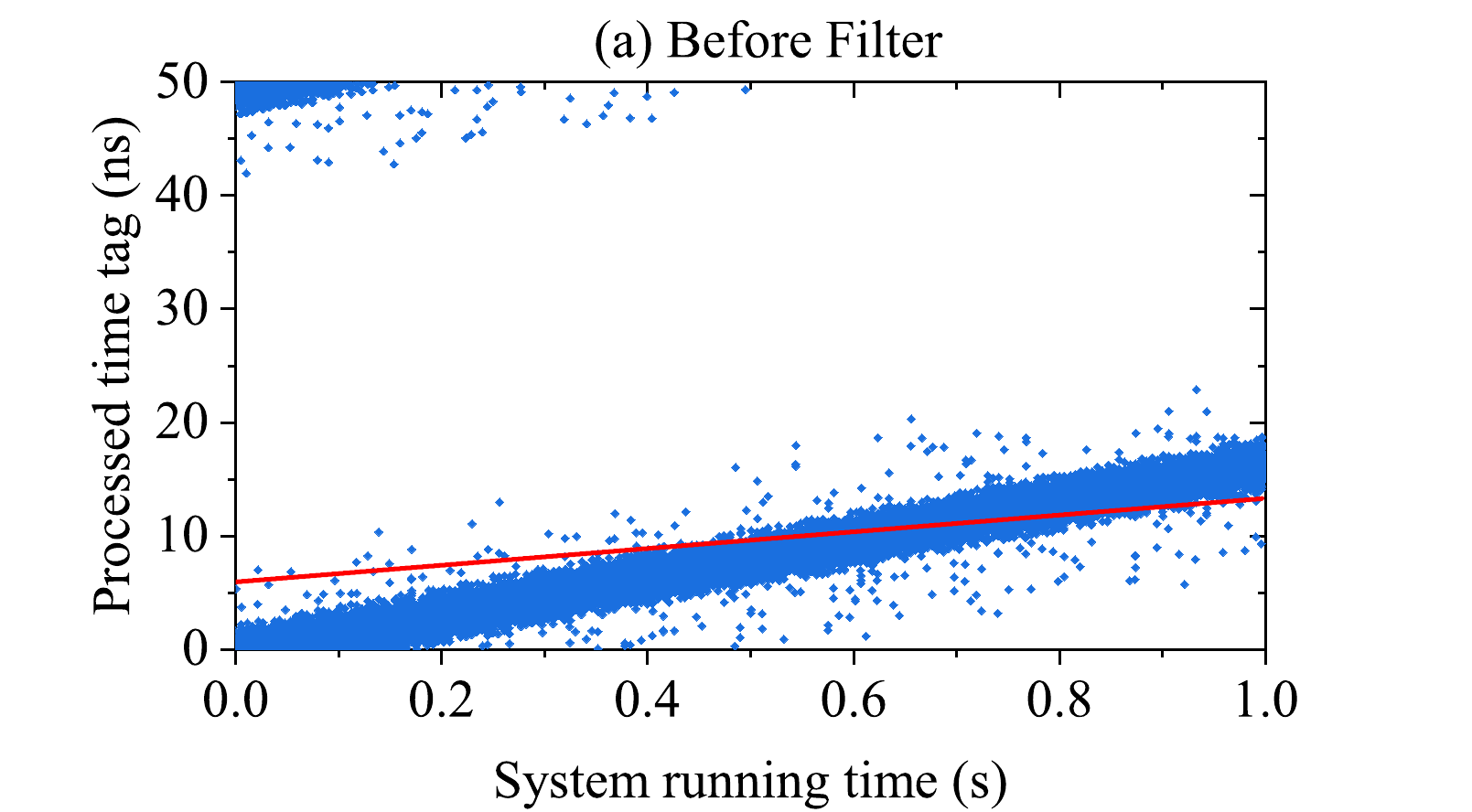 }
	\includegraphics[width=8cm]{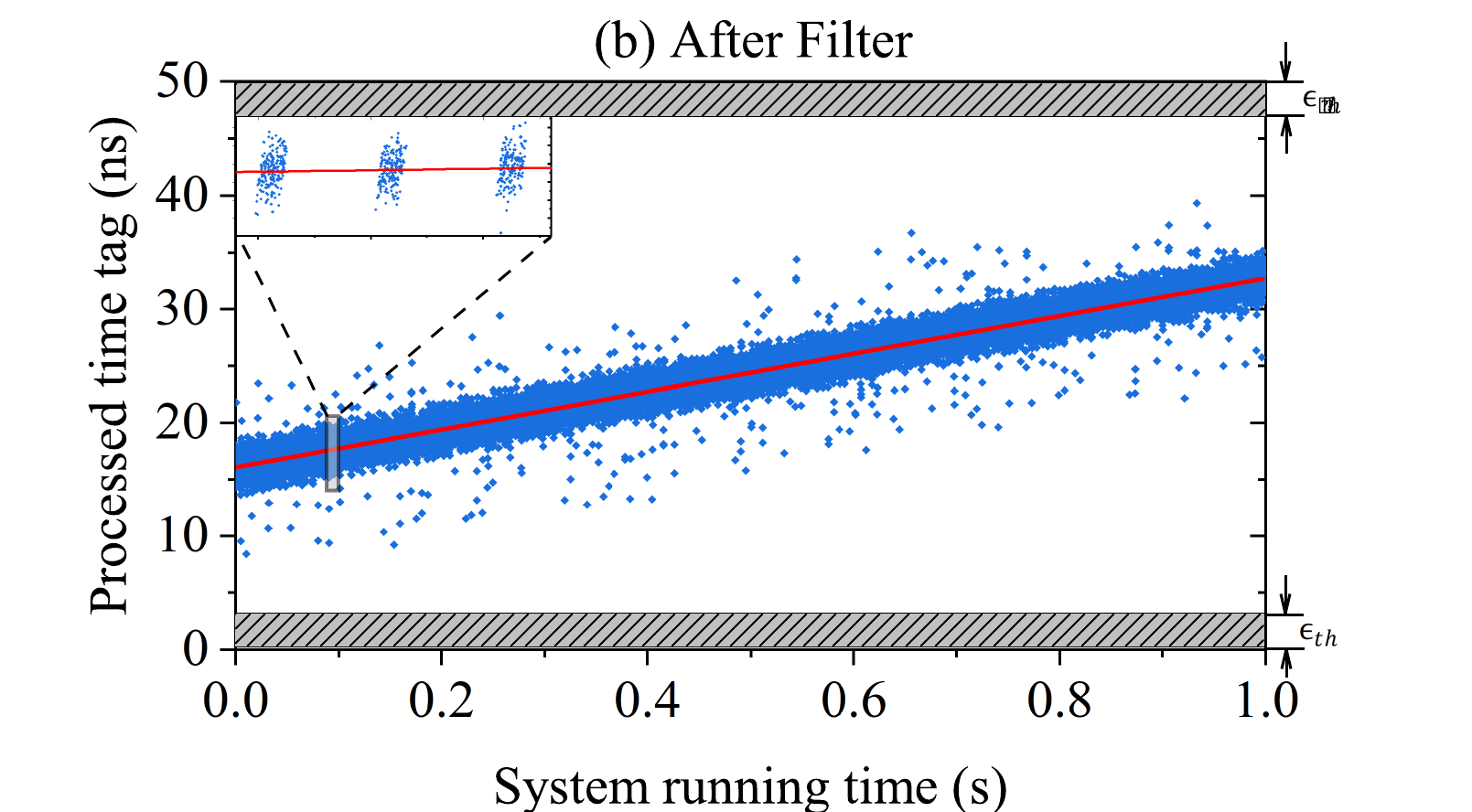 }
	\caption{\label{fig: dataFilter} \textbf{ Monte Carlo simulation result of the data filter module.} The blue dots and red line represent Bob's time data point and the regression results, respectively. The x and y-axis represent the original time tag $\hat{t}_j$ and the processed time tag $y_j$ ($y'_j$). In subfigure (a), the time data are processed by Eq. (\ref{eq 22222}). The result indicates that the previous Qubit4Syncs may be affected by large jitter. In subfigure (b), the time data are processed by Eq. (\ref{eq: new dataset prepross}). Data points falling into the shadow area are filtered according to Eq. (\ref{eq: filtered set}). The result indicates our algorithm robustness against the large jitter. }
\end{figure}

\begin{figure}[htbp]
    \includegraphics[width=8cm]{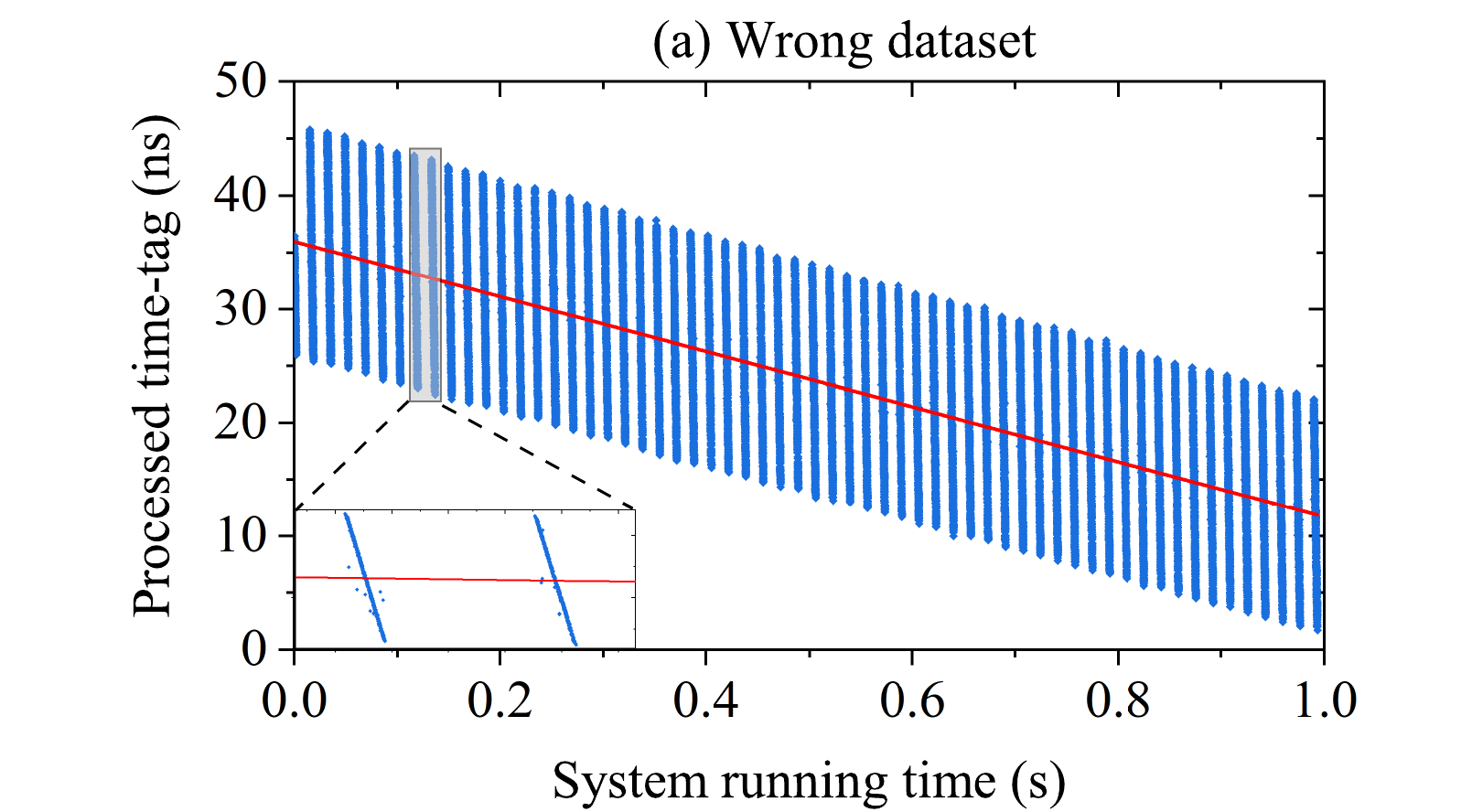 }
        \includegraphics[width=8cm]{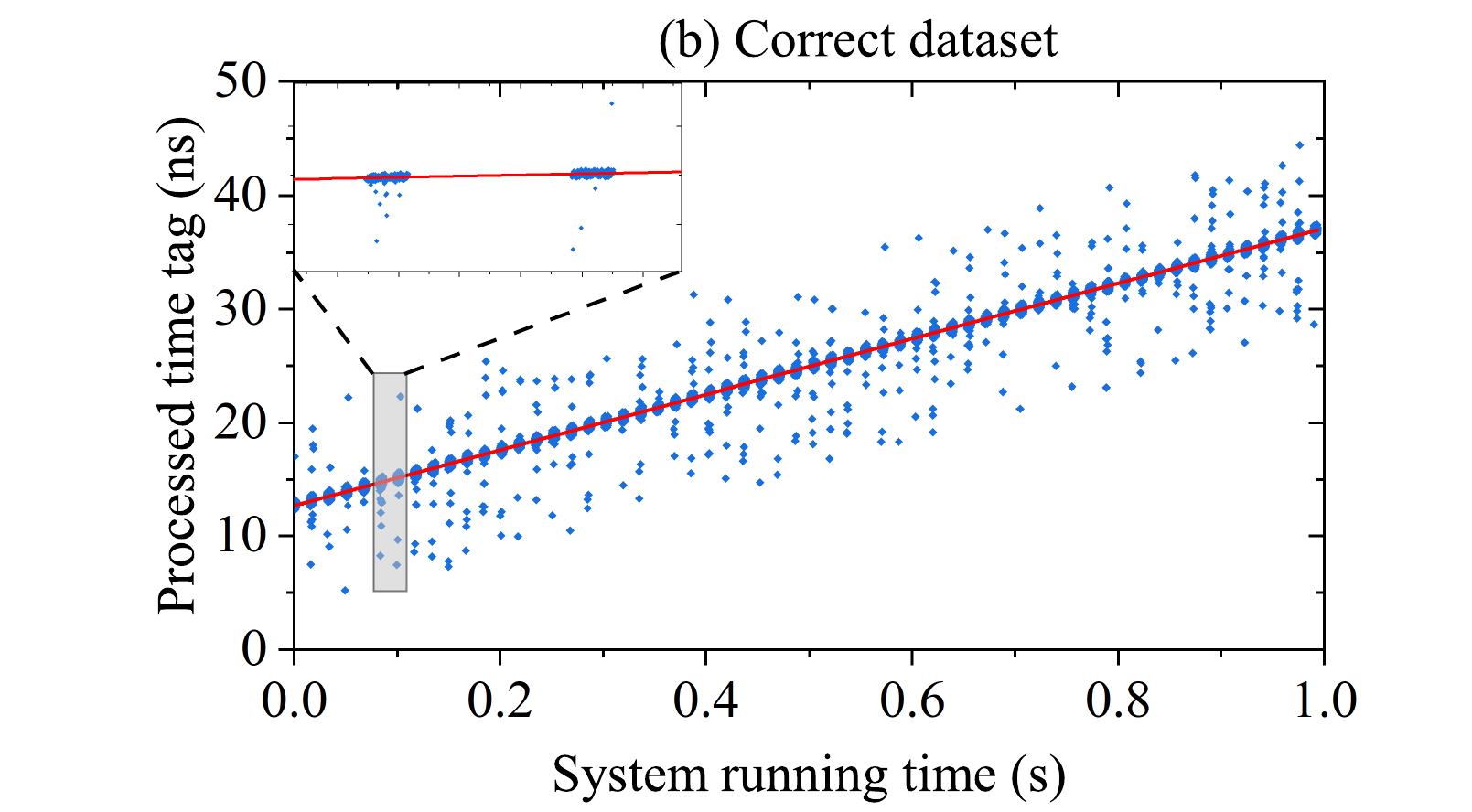}
        \caption{\label{fig: sign-identification} \textbf{Monte Carlo simulation result of the sign-identification module.} Here we suppose positive sign is correct. In subfigure (a), the blue dots represent the dataset $({y'}^-_j, k^-_j)$, which indicates a large $D^-_j$. In subfigure (b), the blue dots represent the dataset $({y'}^+_j, k^+_j)$. The result indicates a small $D^+_j$. }
\end{figure}

  \textbf{Module 3. Sign-identification}:    
    Since $|\hat{\Delta}_{\tau_{coa}}|$ is an absolute value, an additional module is required to judge the sign. 
    Bob employs the data filtering module to prepares a `positive dataset' $\{({y'}^+_j, k^+_j) \big| j \in \mathbb{J}^+ \}$ and a `negative dataset' $\{({y'}^-_j, k^-_j) \big| j \in \mathbb{J}^- \}$, where
 \begin{equation}
\label{eq dataset posi}
\begin{aligned}
        &{y'}^+_j = y'_j \mod (\tau_B + |\hat{\Delta}_{\tau_{coa}}|), \\
        & k^+_j = \lfloor ( t_j - t_0 ) \big/ (\tau_B + |\hat{\Delta}_{\tau_{coa}}|) \rfloor
\end{aligned}
\end{equation}
and 
 \begin{equation}
\label{eq dataset nage}
\begin{aligned}
        &{y'}^-_j = y'_j \mod (\tau_B - |\hat{\Delta}_{\tau_{coa}}|), \\
        & k^-_j = \lfloor ( t_j - t_0 ) \big/ (\tau_B - |\hat{\Delta}_{\tau_{coa}}|) \rfloor
\end{aligned}
\end{equation}
respectively. The two datasets are independently filtered according to Eq. (\ref{eq: filtered set}) and obtain $\mathbb{J}^+$ and $\mathbb{J}^-$ respectively.
Only one dataset yields a valid coarse estimation, which generates a linear regression curve. In contrast, the incorrect one exacerbates the clock misalignment between Alice and Bob. The plot of the two datasets present the distribution as shown in Fig. \ref{fig: sign-identification}. Defining  
 \begin{equation}
\label{eq diff}
\begin{aligned}
D^\pm_j = |{y'}^\pm_{j'} - {y'}^\pm_{j}| \big/ (k^\pm_{j'} - k^\pm_{j}),
\end{aligned}
\end{equation}
where $j,j' \in \mathbb{J}^\pm$ and the two element are adjacent in $\mathbb{J}^\pm$.
It is obvious that the correct $D^\pm_j < |\hat{\Delta}_{\tau_{coa}}| $ and the incorrect $D^\pm_j > \tau_B \gg |\hat{\Delta}_{\tau_{coa}}| $.  To identify the correct dataset, Bob only needs to calculate the average derivatives $\bar{D}^\pm$ and compare their magnitude. The smaller one corresponds to the correct dataset, and the sign of $\Delta_{\tau_{coa}}$ follows the correct dataset. We re-define the correct dataset as $\{({y'}_j, k_j) \big| j \in \mathbb{J} \}$ and the coarse estimated period is $\tau_{coa}= \tau_B + \Delta_{\tau_{coa}}$\\

\textbf{Module 4. Frequency fine-tuning}: 
 Regression algorithms are employed at this stage to enhance precision. Bob estimates his fine-tuning frequency by LSR
 \begin{equation}
\label{eq LSR fin}
\begin{aligned}
  \hat{\tau}_{fin} = \sum_j (k_j - \bar{k}) ( y'_j - \bar{y'} )  \big/ \sum_j (k_j - \bar{k}),
\end{aligned}
\end{equation}
where $\bar{k}$ and $\bar{y'}$ are average values. Bob finally fine-tune his period to $ \tau_B = \hat{f}_{coa} + \hat{f}_{fin}$, which is sufficiently close to $\tau_A$.\\

\textbf{Module 5. Delay tuning }: When the clocks are sufficiently close, Alice and Bob continue operating their systems for a short while, and Bob records all arrival time $\hat{t}_j$. There exist two cases: (1) The SPD yield is stable, and slightly fluctuates near the regular value. This indicates the pulses arrive when the gate is open. (2) The SPD yield decreases to the noise level, which indicates that the pulses are out of the gates. In the first scenario, the users only need to calculate $\bar{y}'$ and fine-tune the SPD-delay to make $\bar{y}' \approx \tau_B / 2$. In the second scenario, Bob should scan the SPD-delay until the first scenario is observed, and then fine-tunes the SPD-delay $\bar{y}' \approx \tau_B / 2$. It is worth noting that \textit{Case (1)} usually only occurs during the initial clock synchronization. When both clocks are already synchronized and tracked in real-time via data, typically only \textit{Case (2)} needs to be considered.\\

When the communication is established, Alice and Bob agree on a common system frequency. However, due to clock inaccuracies, there exists a clock frequency difference $\Delta_f$. As illustrated in Fig. \ref{fig: our sync algorithm}, the synchronization is divided into two stages. The first stage is an initial synchronization. Bob calls the \textbf{Fourier analysis module} and \textbf{sign-identification module} for estimating $\hat{\tau}_{coa}$. After that, the \textbf{frequency fine-tuning module} and \textbf{Delay recovery module} are employed to synchronize their clocks and adjust Bob's SPD-delay to a sufficient precision. Then, the algorithm proceeds to the second stage in which Bob only needs to track Alice's clock drift (without loss of generality, we suppose Bob holds the time reference). In this stage, Bob periodically tracks Alice’s clock using his time tags. He treats his current period as $\hat{\tau}_{coa}$, inputs it into the fine-tuning module to calculate $\hat{\tau}_{fin}$, and subsequently calls \textbf{Delay tuning module} to fine-tune the SPD-delay.

\section{Monte Carlo Simulation}
\label{sec III}

\begin{figure}[htbp]
    \includegraphics[width=7cm]{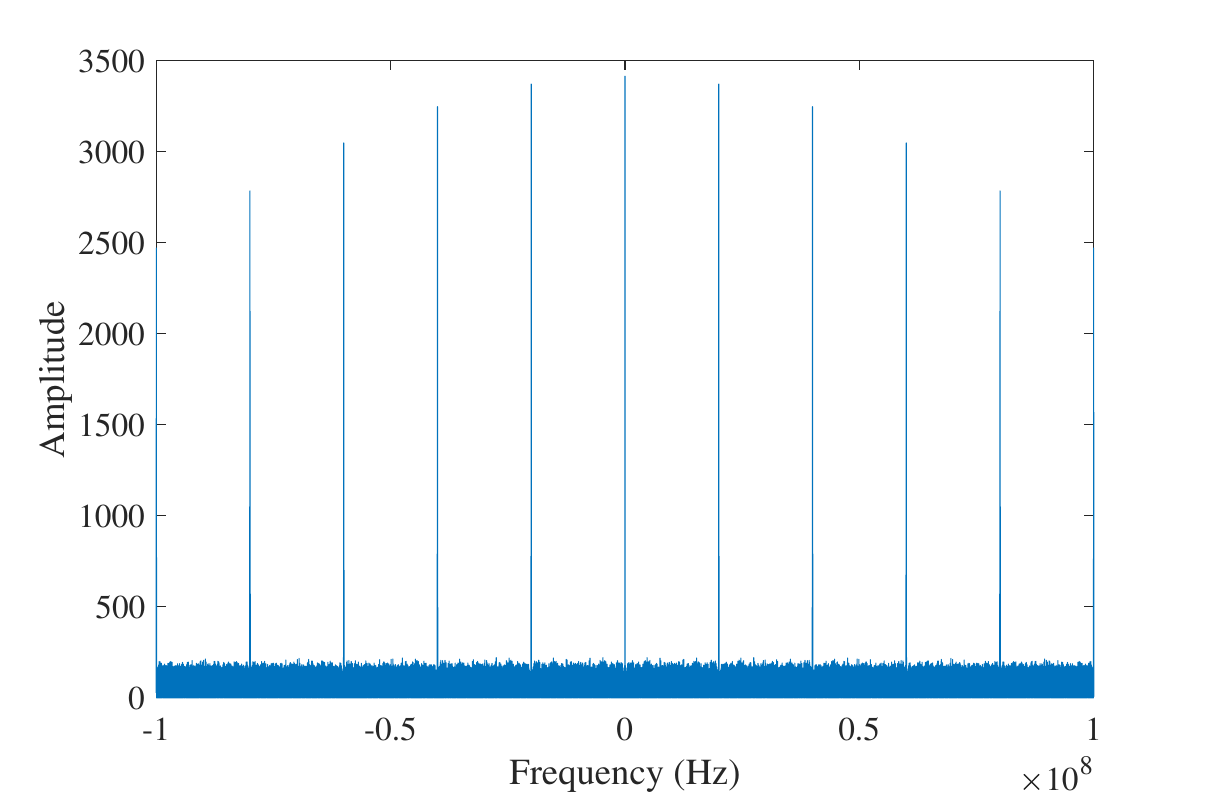 }
        \caption{\label{fig: FFT_ori} \textbf{ Monte Carlo simulation of the Fourier analysis with the previous frequency recovery.} Free-running detection and high-speed sampling are employed. The separation between the two spectral lines is 20.0002 MHz. }
    \includegraphics[width=7cm]{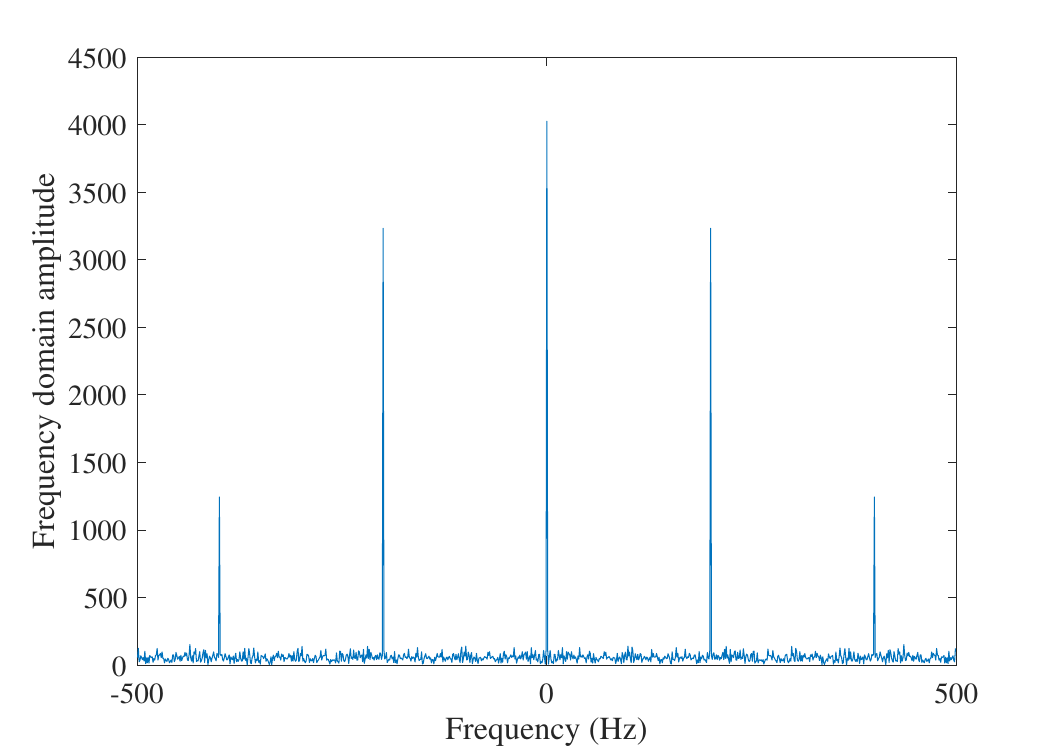 }
    \caption{\label{fig: FFT_imp} \textbf{ Monte Carlo simulation of the Fourier analysis with our frequency recovery.} The separation between the two spectral lines is 200 Hz. The simulation parameters are fixed to 20 MHz system frequency, 1 ns gate width, 200 Hz frequency mismatch, 20 expected \textit{qubit count} per beat, 1 $\rm {\mu s}$ dead time, 50 ps (std) jitter, and 2\% afterpulse rate. }
\end{figure}

To validate the algorithm performance, we employ a Monte Carlo simulation \cite{fan2020universal,huang2024qubit} to mimic a 20 MHz BB84 QKD system with independent and drifting clocks. The simulation considered all regular imperfections in gated-mode detections, Alice employs the decoy-state method \cite{lim2014concise,ma2005practical}, so she randomly selects different intensities, and the weighted mean intensity is 0.1 photon per pulse. Bob employs a gated-mode SPAD with 20\% efficiency and $1$ ns gate width ($T_g$) to detect the pulses. The total jitter is set to 50 ps (std. deviationtion, Gaussian distribution). The dark count rate, afterpulse rate, and dead time are set to $10^{-6}$, 2\%, and 1 $\rm {\mu s}$ respectively. We first generate a 1s ($2\times10^7$ rounds) QKD data with the Monte Carlo simulation in which Alice sends pulses at a 20.0002 MHz repetition rate. In the first case, Bob uses a free-running SPD, employs the previous Qubit4Sync to sample the SPD counts at 200 MHz, and generates $2\times10^8$ sample data. It takes our computer 4.1073 second to process this data by FFT, and 1.47 GB data is finally generated for expressing the frequency spectrum. In contrast, when employing our method, Bob uses a gated mode SPD with 20 MHz gates (or uses free running SPD and post-process it to gate mode). He only needs to sample the SPD counts at 1 kHz and generate 1000 sampling data. Only $1.172\times10^{-4}$ s is consumed to process this data on our computer, and only 20 kB data is finally generated for expressing the frequency spectrum. Compared with the previous methods, our frequency recovery algorithm demonstrates a 35,000 times speedup and a 64,000 times reduction in memory usage in this simulation. The two FFT results are shown in Fig. \ref{fig: FFT_ori} and Fig. \ref{fig: FFT_imp} respectively. 

\hfill

\begin{figure}[htbp]
    \includegraphics[width=8cm]{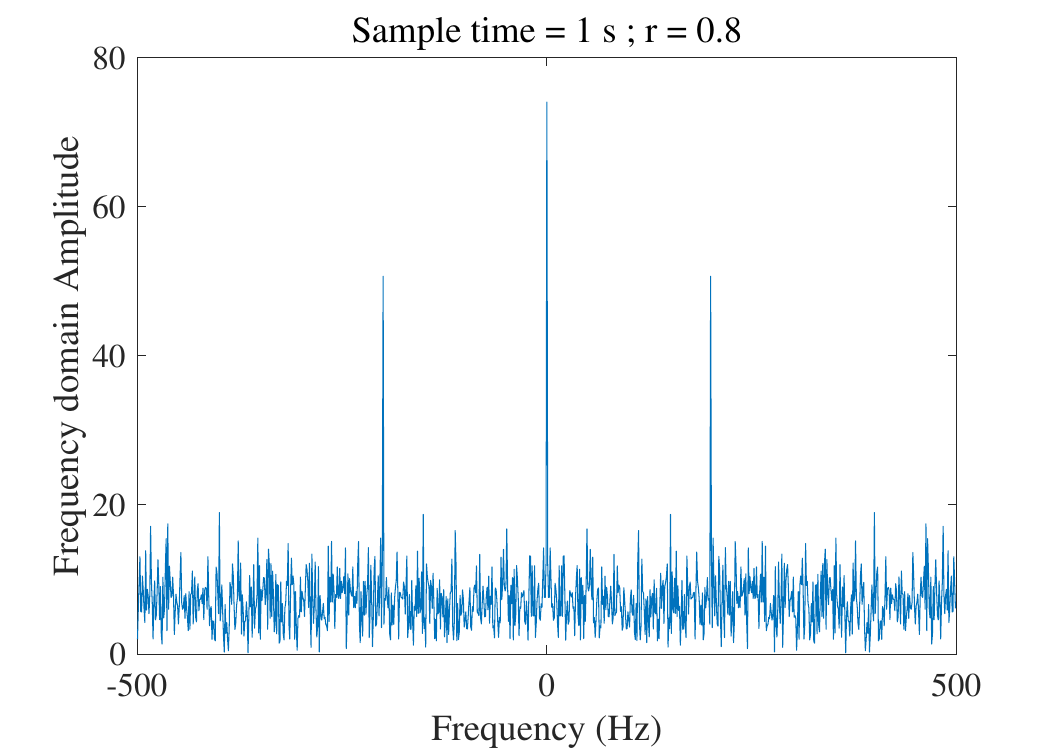 }
    \includegraphics[width=8cm]{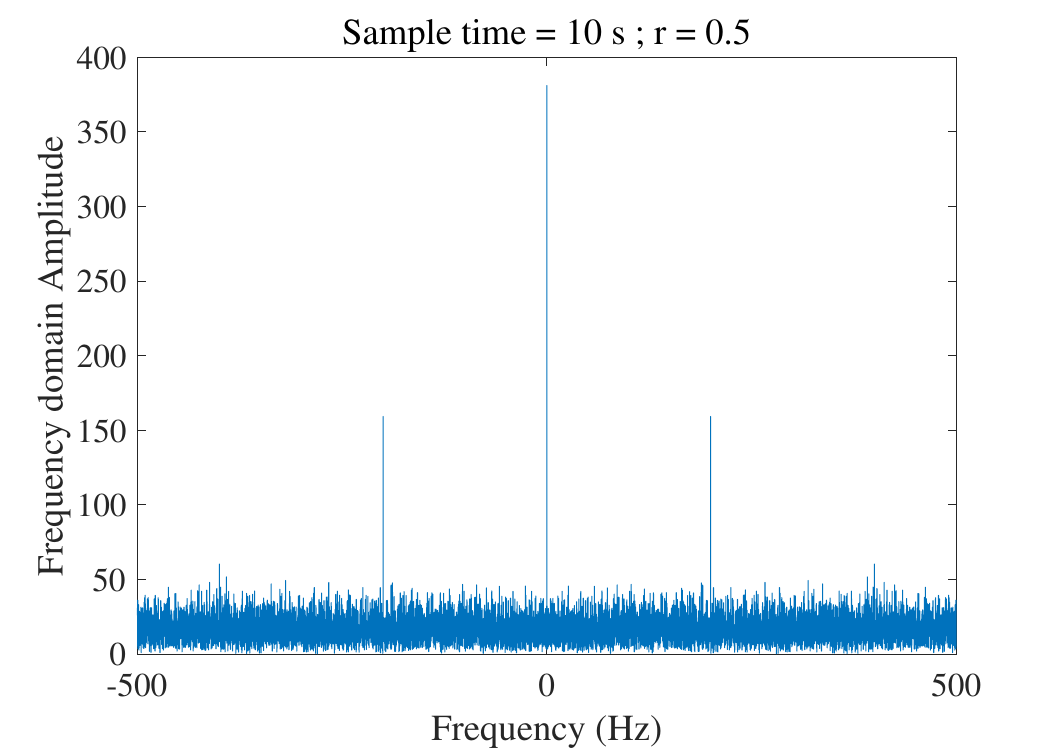 }
    \caption{\label{fig: lowSNR freq} \textbf{ Spectrograms under low SNR conditions. }
    In the first subfigure, the sample time lasts 1 s, consists 80 total qubit counts and 20 dark counts. 
    In the second subfigure, the sample time lasts 10 s, consists 200 total qubit counts and 200 dark counts. 
     Fixed parameters: 20 MHz system frequency, 1 ns gate width, 200 Hz frequency mismatch, 1 $\rm {\mu s}$ dead time, 50 ps (std) jitter, and 2\% afterpulse rate, and $10^{-6}$ dark count rate (about 20 total ).  }
\end{figure}

We also tested the performance of our fast frequency recovery in low SNR scenarios. Here we simulate two scenarios where the ratio of qubit counts to noise counts was 4:1 and 1:1, respectively. The corresponding FFT results are shown in Fig. \ref{fig: lowSNR freq}, demonstrating that the beat frequency peak remains clearly distinguishable even under such low SNR conditions. The detailed theoretical explanation is provided in the Appendix.

\hfill

\begin{figure}[htbp]
    \includegraphics[width=8cm]{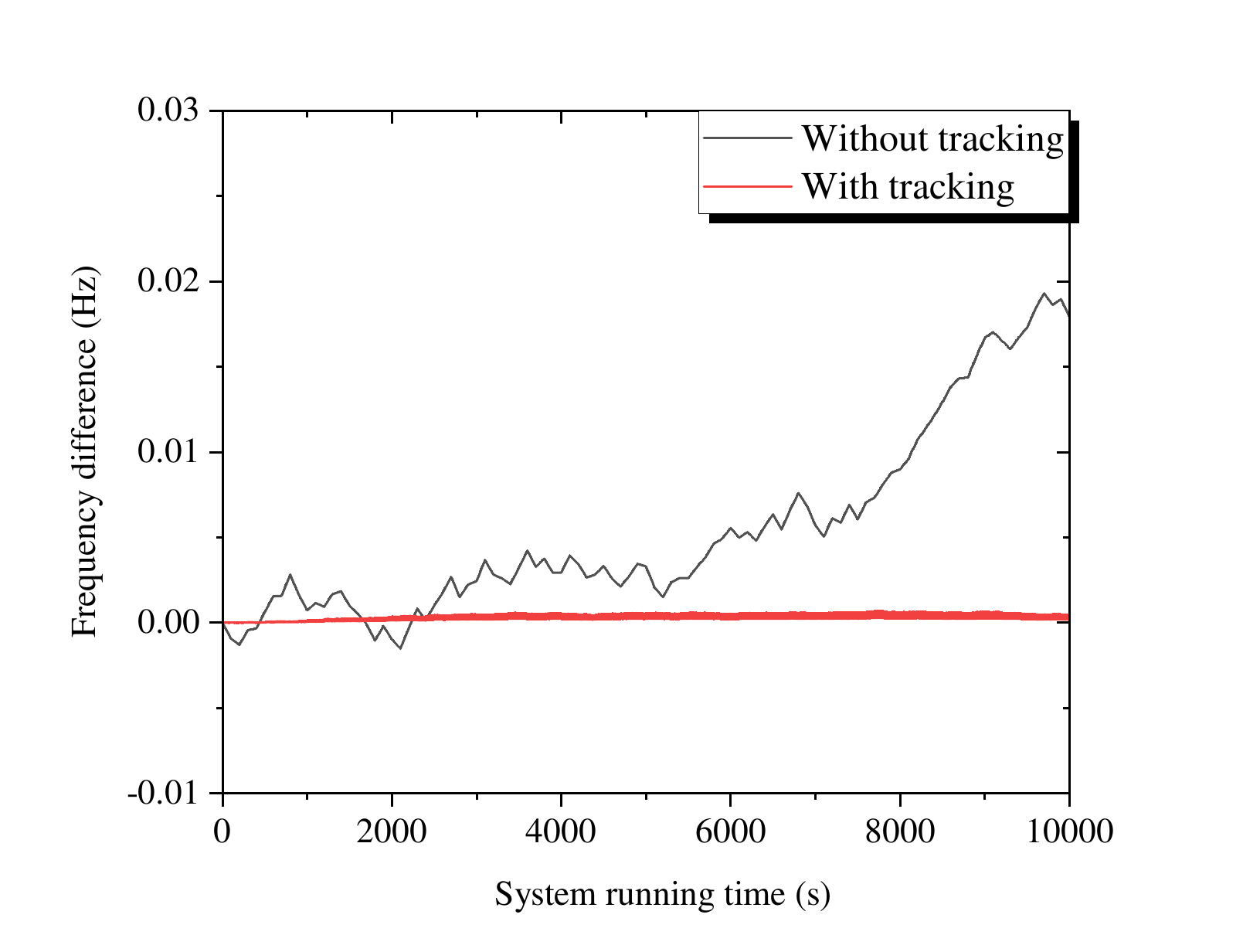 }
        \caption{\label{fig: freq_drift_simu 20M} \textbf{ Monte Carlo simulation of the clock frequency drift at 20 MHz QKD system.}  The x-axis and y-axis represent the system running time and the clock frequency misalignment $\Delta_f$, respectively. The black and red lines represent the system without and with our Qubit4Sync tracking, respectively. The total jitter, dark count rate, afterpulse rate, and dead time are set to 50 ps, $10^{-6}$, 2\%, and 1 $\rm {\mu s}$ respectively.}
             \includegraphics[width=8cm]{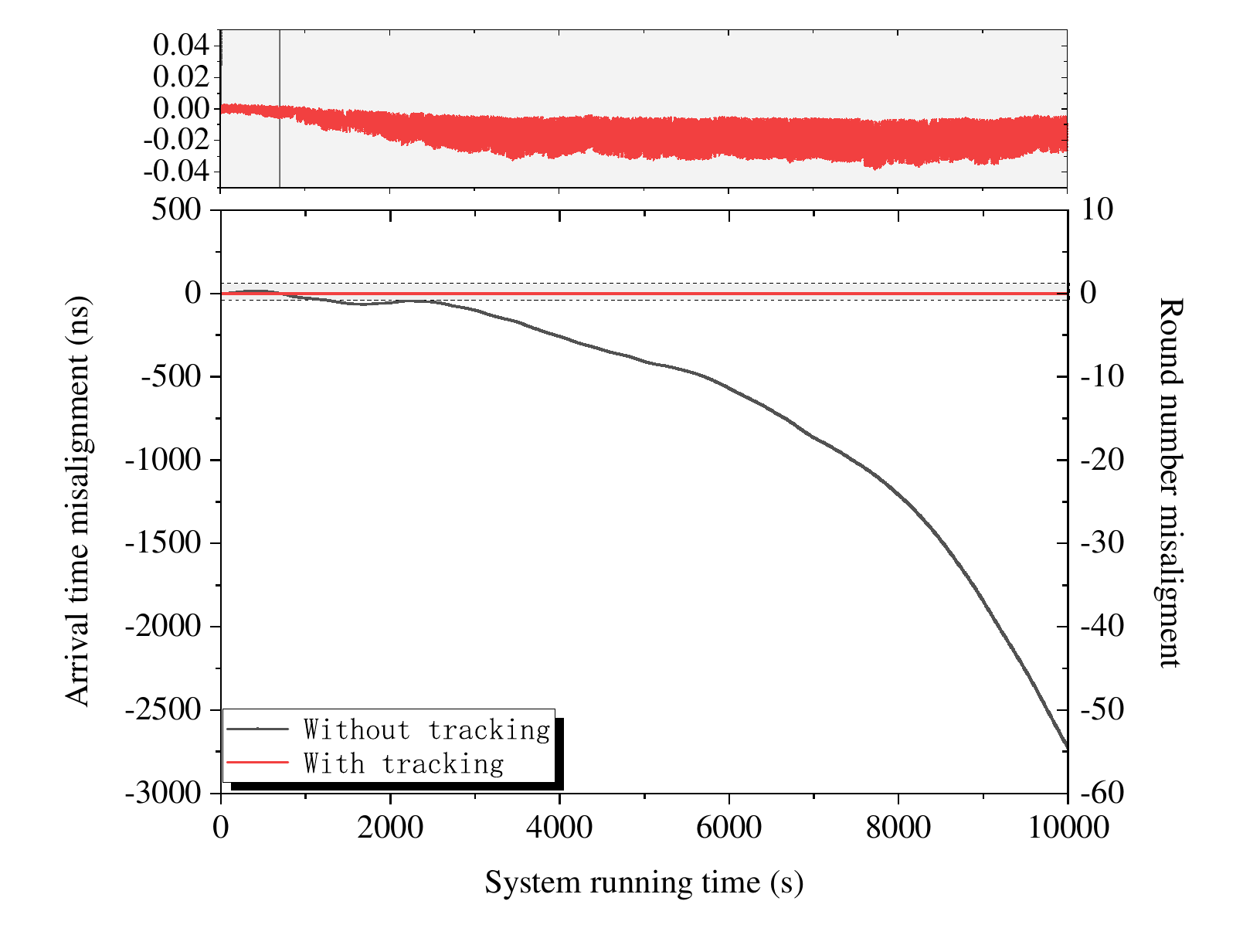 }
  \caption{\label{fig: round_number_drift 20M}  \textbf{ Monte Carlo simulation of the photon arrival time drift at 20 MHz QKD system.} The x-axis and the left y-axis represent the system running time, and the arrival time misalignment, respectively. The right y-axis represents the round number misalignment converted from the 
  time misalignment (50 ns per round). The black and red lines represent the system without and with our Qubit4Sync tracking, respectively. The top subfigure provides an enlarged display of the dashed-box region. }
    \end{figure}

\begin{figure}[htbp]
    \includegraphics[width=8cm]{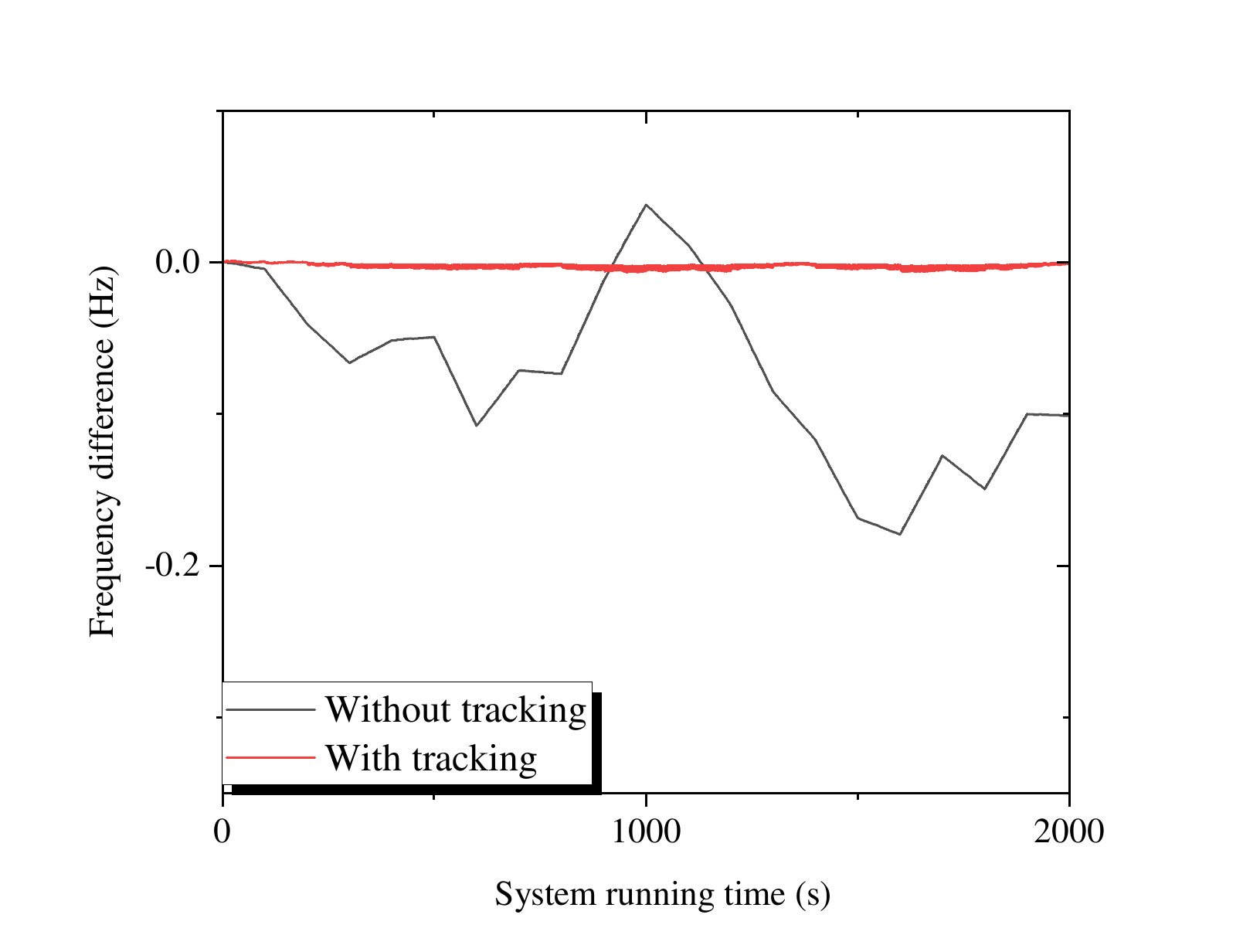 }
        \caption{\label{fig: freq_drift_simu 1G} \textbf{ Monte Carlo simulation of the clock frequency drift at 1 GHz QKD system.}  The x-axis and y-axis represent the system running time and the clock frequency misalignment $\Delta_f$, respectively. The black and red lines represent the system without and with our Qubit4Sync tracking, respectively. The total jitter, dark count rate, afterpulse rate, and dead time are set to 50 ps, $10^{-6}$, 2\%, and 1 $\rm {\mu s}$ respectively.}
             \includegraphics[width=8cm]{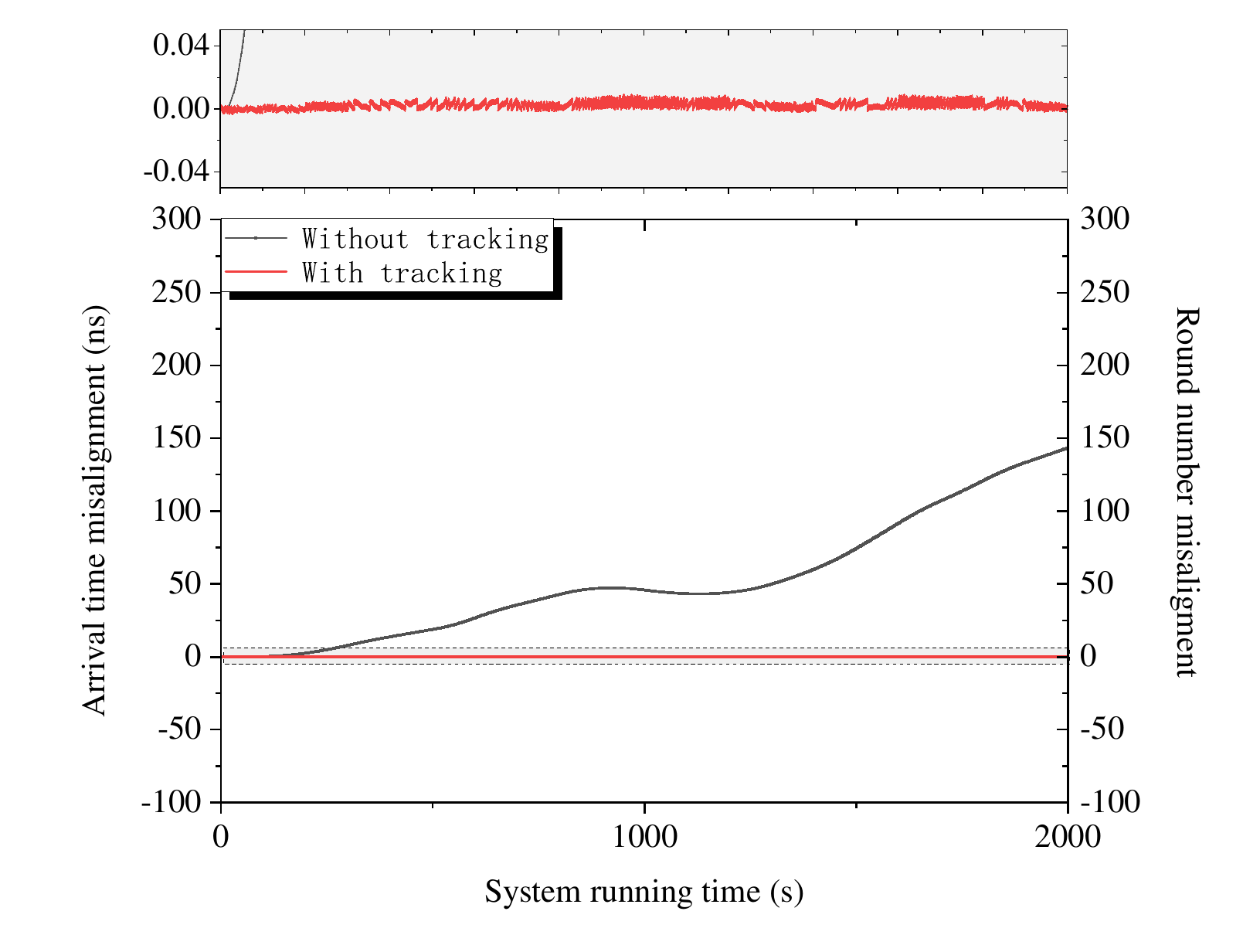 }
  \caption{\label{fig: round_number_drift 1G} \textbf{ Monte Carlo simulation of the photon arrival time drift at 1 GHz QKD system.} The x-axis and the left y-axis represent the system running time, and the arrival time misalignment, respectively. The right y-axis represents the round number misalignment converted from the 
  time misalignment (1 ns per round). The black and red lines represent the system without and with our Qubit4Sync tracking, respectively. The top subfigure provides an enlarged display of the dashed-box region. }
    \end{figure}

Then we validate the clock tracking. Defining $f_A(j)$ ($f_B(j)$) as the Alice's (Bob's) clock frequency at Alice's $j^{\text{th}}$ round, $\hat{t}(j) = \hat{t}(j-1) + 1/f_A(j) + \chi(j) $ as the arrival time of Alice's $j^{\text{th}}$ pulse where $\chi(j)$ is a Gaussian distribution random number for time jitter, $t_g(j)$ as the open-time of Bob's $j^{\text{th}}$ gate, and $\dot{f}_A(j) = f_A(j) - f_A(j-1)$ as Alice's frequency drift velocity. Supposing the initial synchronizing has greatly synchronized, so $f_A(0) = f_B(0) = 2\times 10^7 $ Hz, $t_g(0) = 0$, and $\hat{t}(0) = T_g/2$. Then we define $\dot{f}_A(j) = \dot{f}_{fa}(j) + \dot{f}_{sl}(j)$ where $\dot{f}_{fa}(j)$ and $\dot{f}_{sl}(j)$ are corresponding fast drift component and slow drift component respectively. In the simulation, we update $\dot{f}_{fa}(j)$ and $\dot{f}_{sl}(j)$ by two different random number generators. The random numbers following uniform distribution and satisfies $ Var(\dot{f}_{sl}(j))  \ll Var(\dot{f}_{fa}(j)) $.
Based on $\dot{f}_A(j)$ we further deduce $f_A(j)$ and $\hat{t}(j)$. Based on the above definitions, a 20 MHz system and an 1 GHz system are simulated two different cases are simulated. In each of the systems, we simulate two different case. In the first case, Bob does not track Alice's clock so $t_g(j) = j / f_B(j)$ ($f_B(j) \equiv 2\times 10^7$ in this case). If $\hat{t}(j)$ belongs to any gate, Monte Carlo \cite{fan2020universal,huang2024qubit} is employed to calculate Bob's time tag. In the second case, Bob tracks Alice's clock through our Qubit4Sync with a 2 s interval ($j \mod 4\times 10^7 = 0$). We still simulate Bob's time tags as in the first case, but Bob exploits his time tags to execute the Qubit4Sync for updating $f_B(j)$ and adjusting the gate delay. The simulation results are shown in following four Figures. Figure \ref{fig: freq_drift_simu 20M} (Fig. \ref{fig: freq_drift_simu 1G}) illustrates the frequency misalignment $\Delta_f$ at the 20 MHZ (1 GHz) systems. The Qubit4Sync locks the clock misalignment at the $10^{-4}$ ($10^{-3}$) Hz level. In contrast, without actively tracking, the clock misalignment {reaches 0.02 (0.2) Hz level}. This implies that Alice and Bob accumulate one additional round of misalignment every 50 (5) seconds. Worse still, the pulse can exit the gate region on the ms timescale, which is sufficient to crash the system. Figure \ref{fig: round_number_drift 20M} (\ref{fig: round_number_drift 1G}) illustrates the corresponding round misalignment at 20 MHz (1 GHz) system. When Qubit4Sync is employed, no round misalignment occurred throughout the nearly 3-hour (2000 s) system operation. In contrast, without frequency tracking, the accumulated arrival time misalignment reaches 3000 (150) ns, equals to 60 (150) rounds. In contrast, when tracking is employed, the arrival time misalignment is locked in $\pm 40$ ($\pm 5$) ps level.

\hfill

\section{Conclusion}

In this work, we have advanced the practical deployment of Qubit4Sync by overcoming key limitations of prior approaches. We introduced a fast frequency recovery algorithm that greatly reduces both the data volume and computational time required for clock synchronization, while maintaining robust performance under challenging conditions including low SNR and imperfect detections. Crucially, our method enables reliable operation in mainstream gated-mode QKD systems. Furthermore, we established a comprehensive theoretical framework that incorporates non-ideal detector effects, provides explicit frequency-domain SNR estimation, and offers practical guidance for parameter selection in real-world implementations. Extensive Monte Carlo simulations across 20 MHz and 1 GHz systems validate the efficiency and accuracy of our approach. By simultaneously addressing the algorithmic and theoretical gaps in Qubit4Sync, this study not only enhances its applicability across diverse QKD platforms but also removes obstacles toward hardware-independent synchronization, paving the way for more scalable, cost-effective, and field-ready quantum networks.


\begin{acknowledgments}
This work was supported in part by the National Natural Science Foundation of China under Grant 62301524, Grant 62425507, Grant 62271463, Grant 62105318, Grant 61961136004, and Grant 62171424, in part by the Fundamental Research Funds for the Central Universities, and in part by China Postdoctoral
Science Foundation under Grant 2022M723064, in part by the Natural Science Foundation of Anhui under Grant 2308085QF216, and in part by the Innovation Program for Quantum Science and Technology under Grant 2021ZD0300700. 

We thank Dr. Jie Wu for insightful discussions and guidance on statistics.
\end{acknowledgments}

\hfill

\appendix

\section{Theoretical foundation for qubit-based synchronizations}
\label{sec IV}

To address the theoretical gaps in Qubit4Sync and provide practical guidance for algorithm parameter designing, we propose a theoretical model for the frequency recovery in this section. This model characterizes the impact of the non-ideal properties on the performance of frequency recovery and quantitatively describes their influence on the SNR ratio and synchronization accuracy.

Before introducing the theory, we model the QKD process as follows. In each round, Alice randomly prepares a quantum state $\ket{r}\in \mathbb{R}$ from a pre-decided encoding setting $\mathbb{R}$. If the decoy-state method \cite{hwang2003quantum,wang2005beating,lo2005decoy} is employed, she should also randomly modulate intensity $a \in \mathbb{A}$ where $\mathbb{A}$ is a pre-decided intensity setting. Alice sends her quantum state to Bob through a public channel. Bob randomly selects a basis $b$ from his pre-decided basis setting $\mathbb{B}$ to measure the quantum state. We define $s = \{ r, a, b \} $ as the combination of the selections and $s_k$ as the selection combination of the $k^{\text{th}}$ round.  The probability for selecting $s$ is $P_s$. Let $N_s$ be the number of rounds where setting $s$ is selected, and $n_s$ be the corresponding number of rounds in which an SPD count occurs. Then the count rate for setting $s$ is given by $Q_s = n_s / N_s$. The average count rate of the SPD is $Q = \sum_{s} P_s Q_s$.




\subsection{Ideal free-running detection}

For simplicity, we first model an ideal free-running detection. In this scenario, Alice periodically sends her quantum state to Bob via a lossy channel with period $\tau_A$, Bob detects the quantum states with his free-running SPDs. 

Define the successful event at time $\hat{t}$ as the Dirac delta function $\delta(t - \hat{t})$, we can express the successful events in the time domain as
\begin{equation}
\label{eq 1-1}
x_1(t) = \sum_{k \in \mathbb{Z}} c_k\delta(t-k\tau_A),
\end{equation}   
where $c_k\sim Bernoulli(Q_{s_k})$ are random variables following a Bernoulli distribution with parameter $Q_{s_k}$. $c_k = 1$ if Bob observes a successful event and $c_k = 0$ if failure. Then, the frequency domain is expressed as
\begin{equation}
    \label{eq 1-2}
    {X}_1(f) = \sum_{k \in \mathbb{Z}} c_k e^{-i 2\pi f k\tau_A},
\end{equation}  
Noting that 
\begin{equation}
    \label{eq 1-2-1}
     \sum_{k \in \mathbb{Z}} e^{-i 2\pi f k\tau_A} = f_A\sum_{n \in \mathbb{Z}} \delta\left(f -n f_A \right),
\end{equation} 
we obtain the expected frequency domain function as 
\begin{equation}
    \label{eq 1-3}
    \langle {X}_1(f) \rangle = f_A Q \sum_{m \in \mathbb{Z}} \delta(f - m f_A),
\end{equation}  
where $\langle \cdot \rangle$ is defined as the expected value of an observable $\cdot$. Equation (\ref{eq 1-3}) indicates that the frequency spectrum of the successful events consists of discrete spectral lines spaced by $f_A$.

\subsection{Impact of the unexpected counts}

To advance the practical deployment of the qubit synchronization, we further account for non-ideal characteristics. In this subsection, we consider the detector flaws, including afterpulse \cite{humer2015simple,he2017sine,fan2018afterpulse}, dark count \cite{kang2003dark}, and dead time \cite{weier2011quantum,huang2022dependency} – all of which introduce additional frequency components in the spectral domain.

We first model the dead time of SPDs. The dead time is prevalent in nearly all practical SPDs, manifesting as a duration $T_d$ during which the detector remains non-responsive following a triggering event. Defining $T_d =: \tau_B {N_d} $ where ${N_d}$ represents the unresponsive round number. With this model, the count function of the SPD still satisfies the form of
\begin{equation}
    \label{eq 2 1 1}
    x_2(t) = \sum_{k \in \mathbb{Z}} \tilde{c}_k \left[ \delta(t-k\tau_A)  \right],
\end{equation} 
but the random variable $\tilde{c}_k$ does not follow the Bernoulli distribution anymore. It correlates to the previous ${N_d}$ rounds (for simplicity, we suppose ${N_d}$ is an integer) and follows a distribution   
\begin{equation}
    \label{eq 2 1 2}
    \tilde{c}_k\mid \mathbf{c}_{k-1}^{({N_d})} \sim 
\begin{cases}
    \text{Binomial}(1, Q), & \text{if } \mathbf{c}_{k-1}^{({N_d})} = \mathbf{0}, \\
    0, & \text{otherwise},
\end{cases}
\end{equation} 
where $\mathbf{c}_{k-1}^{({N_d})} = ( c_{k-{N_d}}, c_{k-{N_d}+1}, \dots, c_{k-1} )$ represents the SPD-click of the previous ${N_d}$ rounds.
Based on the previous our \cite{huang2022dependency}, we get 
\begin{equation}
    \label{eq 2 1 3}
    \langle \tilde{c}_k \rangle = \frac{\langle c_k \rangle}{1 + T_d\langle c_k \rangle}  = \frac{Q}{1 + T_d Q},
\end{equation} 
which indicates that the dead time does not introduce any additional frequency components and only affects the energy of the frequency domain. 

Then we add the afterpulse to our model: given a SPD count at time $t'$, there exists a probability $P_{ap}$ for an afterpulse to occur within temporal interval $[ t' + T_d, +\infty ]$. The count function is expressed as
\begin{equation}
    \label{eq 2 2 1}
\begin{aligned}
        x_3(t) =& \sum_{k \in \mathbb{Z}} \tilde{c}_k [ \delta(t-k\tau_A) +\\& \hat{c}_k \delta(t-k\tau_A-T_d-\epsilon_k) ],
\end{aligned}
\end{equation} 
where $\hat{c}_k\sim Bernoulli(P_{ap})$ is a random variable, and $\hat{c}_k$ indicates whether the count of the $k^\text{th}$ round triggers an afterpulse in the following rounds or not. $k\tau_A+T_d+ \epsilon_k$ denotes the afterpulse triggered by the count of the $k^\text{th}$ round, $\epsilon_k \sim Exp(\lambda)$ is also a random variable following the exponential distribution whose probability density function is expressed as $P(\epsilon_k) = \lambda e^{-\lambda \epsilon_k}$ for $\epsilon_k \geq 0$. Besides, considering the afterpulse, the expected value of $\tilde{c}_k$ is revised as 
\begin{equation}
    \label{eq 2 2 2}
\begin{aligned}
    \langle \tilde{c}_k \rangle &= \frac{\langle c_k \rangle}{1 + T_d\langle c_k \rangle (1+\langle \hat{c}_k \rangle)} \\& = \frac{Q}{1 + T_d Q(1+P_{ap})} := \tilde{Q}
\end{aligned}
\end{equation} 
We emphasize that Eq. (\ref{eq 2 2 2}) is not a rigorous result, as we have neglected certain higher-order small quantities \cite{huang2022dependency}. With the above definitions, the expected frequency domain function satisfies
\begin{equation}
    \label{eq 2 2 3} 
\begin{aligned}
    &\langle {X}_3(f) \rangle =  \sum_{k \in \mathbb{Z}}
      \left[ \langle \tilde{c}_k \rangle e^{-i2\pi f k\tau_A} + \langle  \tilde{c}_k \hat{c}_k \rangle  \langle e^{-i 2\pi f (k\tau_A + T_d + \epsilon_n)}  \rangle  \right]
      \\ & = f_A \sum_{m \in \mathbb{Z}}  \delta(f - m f_A)
      \left[  \left(\tilde{Q} + \tilde{Q}_{ap}\frac{  \lambda}{i 2\pi f + \lambda} e^{-i 2\pi f T_d}\right)   \right],
\end{aligned}
\end{equation}  
where $ \tilde{Q}_{ap}=:\langle \tilde{c}_k \hat{c}_k \rangle $. Equation (\ref{eq 2 2 3}) indicates that afterpulses do not introduce any additional frequency components, although they count randomly. The afterpulse manifests as an attenuation $\lambda\tilde{Q}_{ap} \big/ (i 2\pi f + \lambda) $ of high-frequency components. 

Finally, we simply model the dark count: obviously, the expected value of the dark count is an addition DC-signal which only affects the 0-frequency component. Apart from this, the probability of dark counts is a higher-order small quantity compared to other counts and thus will not be discussed in the subsequent analysis.

\subsection{Impact of the timing jitter}

Timing jitter is widely existing in all practical QKD systems, arising from laser emission fluctuation, click-time fluctuation, and electronic noise. This makes it a non-negligible factor in theoretical analysis. Here, we model the time jitter and analyze its impact on the qubit-based synchronization. To enhance clarity in exposition, we initially neglect the afterpulse and the dead time. The time jitter breaks the period mode in Bob's record, and we model it as
\begin{equation}
    \label{eq 3 1 1}
    x'_4(t) = \sum_{k \in \mathbb{Z}} \tilde{c}_k \left[ \delta(t-k\tau_A - \chi_n)  \right],
\end{equation} 
where $\chi_k$ follows a Gaussian distribution with 0 mean and $\sigma$ standard deviation, denoted as $\chi_k \sim \mathcal{N}(0, \sigma^2)$. Its expected frequency domain is expressed as
\begin{equation}
    \label{eq 3 1 2}
    \begin{aligned}
    &\langle {X}'_4( f) \rangle = 
    \sum_{k \in \mathbb{Z}} \langle \tilde{c}_k \rangle 
    e^{-i 2\pi f k\tau_A}  \langle e^{-i 2\pi f \chi_k} \rangle\\ & = 
      Q e^{-2{(\pi f \sigma)^2}}  f_A \sum_{m \in \mathbb{Z}} \delta(f - m f_A),
    \end{aligned}
\end{equation} 
where $\langle e^{-i 2\pi \chi_k} \rangle = e^{-2(\pi f \sigma)^2} $ is derived by Cauchy integral theorem. Equation (\ref{eq 3 1 2}) reveals a counterintuitive result: while timing jitter disrupts periodic temporal mode, it does not introduce any additional frequency components in the spectral domain – only an attenuation of high-frequency elements that scales with jitter amplitude. Adding the afterpulse and deadtime, Eq. (\ref{eq 3 1 1}) and Eq. (\ref{eq 3 1 2}) are revised as 
\begin{equation}
    \label{eq f_4}
\begin{aligned}
    &x_4(t) = \sum_{k \in \mathbb{Z}} \tilde{c}_k \big[ \delta(t- k\tau_A - \chi_k) \\&
    + \hat{c}_k\delta(t- k\tau_A - \chi_k - T_d - \epsilon_k) \big],
\end{aligned}
\end{equation}
and 
\begin{equation}
    \label{eq expected F_4}
    \begin{aligned}
\langle {X}_4(f) \rangle  = f_A H(f)  \sum_{m \in \mathbb{Z}} \delta(f - m f_A), 
    \end{aligned}
\end{equation}
where 
$ H(f) = \Big[ \tilde{Q}'   
        + \tilde{Q}'_{ap} \frac{ \lambda}{i 2\pi f + \lambda} e^{-i 2\pi f T_d}  \Big]   e^{-2(\pi f \sigma)^2}. $

\begin{figure}[htbp]
    \includegraphics[width=8cm]{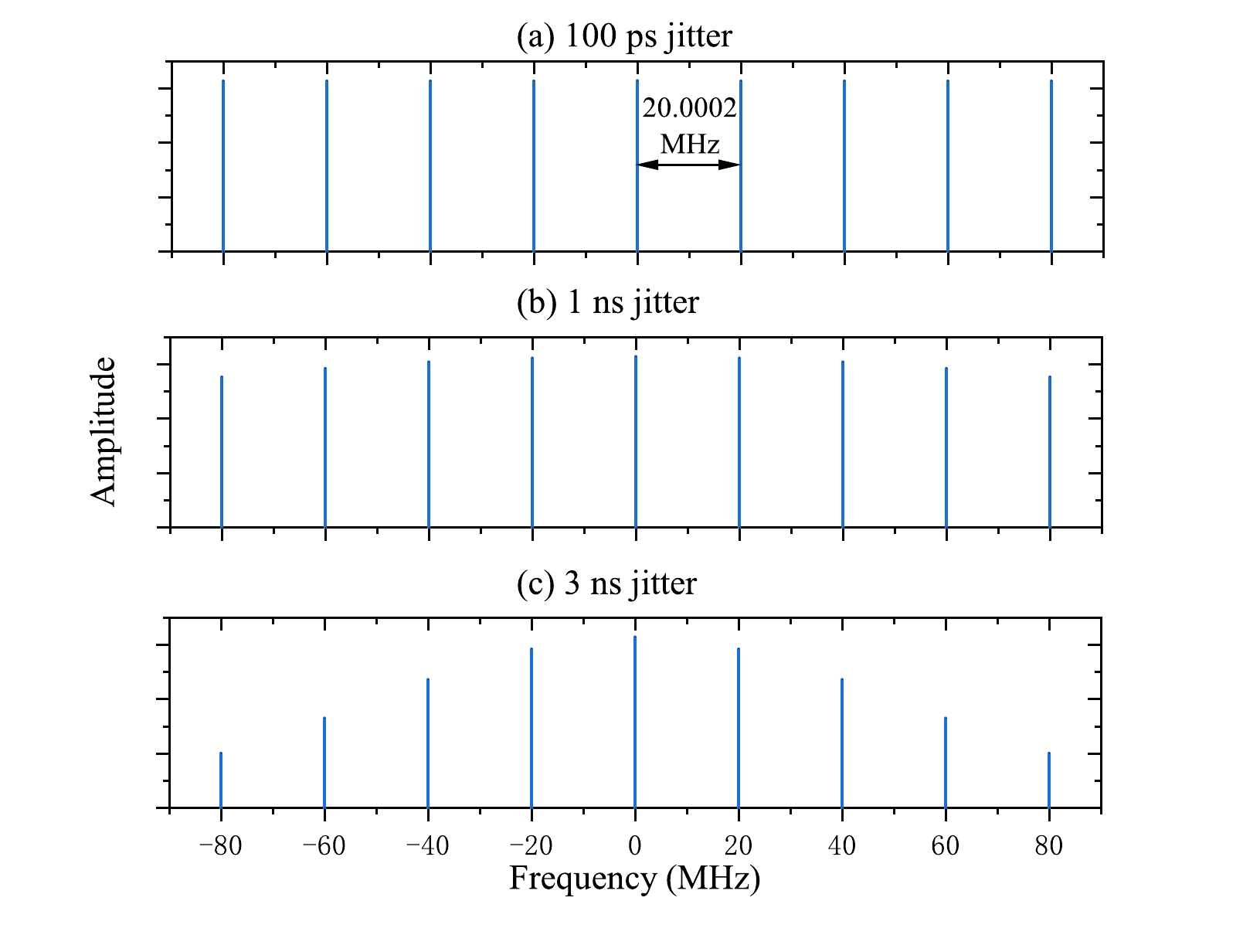 }
    \caption{\label{fig: FFT_jitter} \textbf{ Expected frequency spectrum of a free-running system with time jitter. } Regarding Bob's clock as the reference, Alice prepares and sends the quantum state at 20.0002 MHz. Bob can track Alice's clock by reading the distance of the spectral lines. Subfigures (a), (b), and (c) show that the standard deviations of the total timing jitter in Bob's records are 100 ps, 1 ns, and 3 ns, respectively. Clearly, larger jitter introduces more attenuation in the high-frequency components.}
\end{figure}

\subsection{gated-mode SPD and misaligned clock}

Since gated-mode SPDs still dominate deployed QKDs, understanding and quantifying the misalignment of the clock frequency is essential. Here, we model the impact as follows. The gating process is formally equivalent to a periodic temporal filtering function $ g(t) = \sum_{k \in \mathbb{Z}} g_s(t - k\tau_B) $ where $\tau_B$ is Bob's system period and $g_s(t)$ is a single gate. Without loss of generality, we suppose $g_s(t) = \text{rect}(t - k\tau_B)$, where 
\begin{equation}  
        \text{rect}(t) =
        \begin{cases}
            1, & \text{if } -T_g/2<t<T_g/2, \\
            0, & \text{otherwise},
        \end{cases} 
\end{equation}
and $T_g$ is the gate width. Bob's clicks at the gated mode are expressed as $f_{g} = x_4(t)g(t)$. According to the frequency-domain convolution theorem, we get the frequency domain function as
\begin{equation}  
    \label{eq 4 1}
    {X}_g(f)  =  {X}_4(f)*{G}(f),
\end{equation}
where $  {G}(f) =  f_B T_g  \sum_{m \in \mathbb{Z}}  \text{sinc}\left( m \pi f_B T \right) \delta(f - mf_B) $ is the frequency domain function of Bob's gate, where $f_B = 1 \big/ \tau_B$ is Bob's clock frequency. Noting the sifting property of the Dirac delta function, the expected value of Eq. (\ref{eq 4 1}) is
\begin{equation}
    \label{eq F}
    \begin{aligned}
        &\langle {X}_g(f) \rangle = 
          T_g f_B \sum_{n \in \mathbb{Z}}  \text{sinc}\left( m \pi f_B T_g \right) \langle {X}_4(f - m f_A) \rangle
         \\ & =  T_g f_B f_A \sum_{l \in \mathbb{Z}} \sum_{m \in \mathbb{Z}}  \text{sinc}\left( l\pi f_B T_g \right) 
         \\ & \times H(f - lf_A) \delta(f-l f_B - m f_A), 
    \end{aligned}
\end{equation}
As $\Delta_f = f_A - f_B$ is the clock misalignment between Alice and Bob. We can revise Eq. (\ref{eq F}) as 
\begin{equation}
    \label{eq F revised}
    \begin{aligned}
      & \langle {X}_g(f) \rangle =  T_g f_Af_B \sum_{l \in \mathbb{Z}} \sum_{m \in \mathbb{Z}}  \text{sinc}\left( l \pi f_B T_g \right) \times
      \\& H(f- l f_A) \delta[f - (l+m)f_A + l\Delta_f]. 
    \end{aligned}
\end{equation}
Equation (\ref{eq F revised}) reveals that the frequency spectrum of a misaligned gated-mode system consists of multiple spectral line groups. All lines satisfying $l+m=k$ form a group-$k$, where inter-group distance indicates $f_A$ while intra-group distance is $|\Delta_f|$. Intuitively, this intra-group distance reflects the beat frequency component. It indicates that the users can judge their clock misalignment by measuring the frequency difference between two intra-group spectrum lines. 

\begin{figure}[htbp]
    \includegraphics[width=8cm]{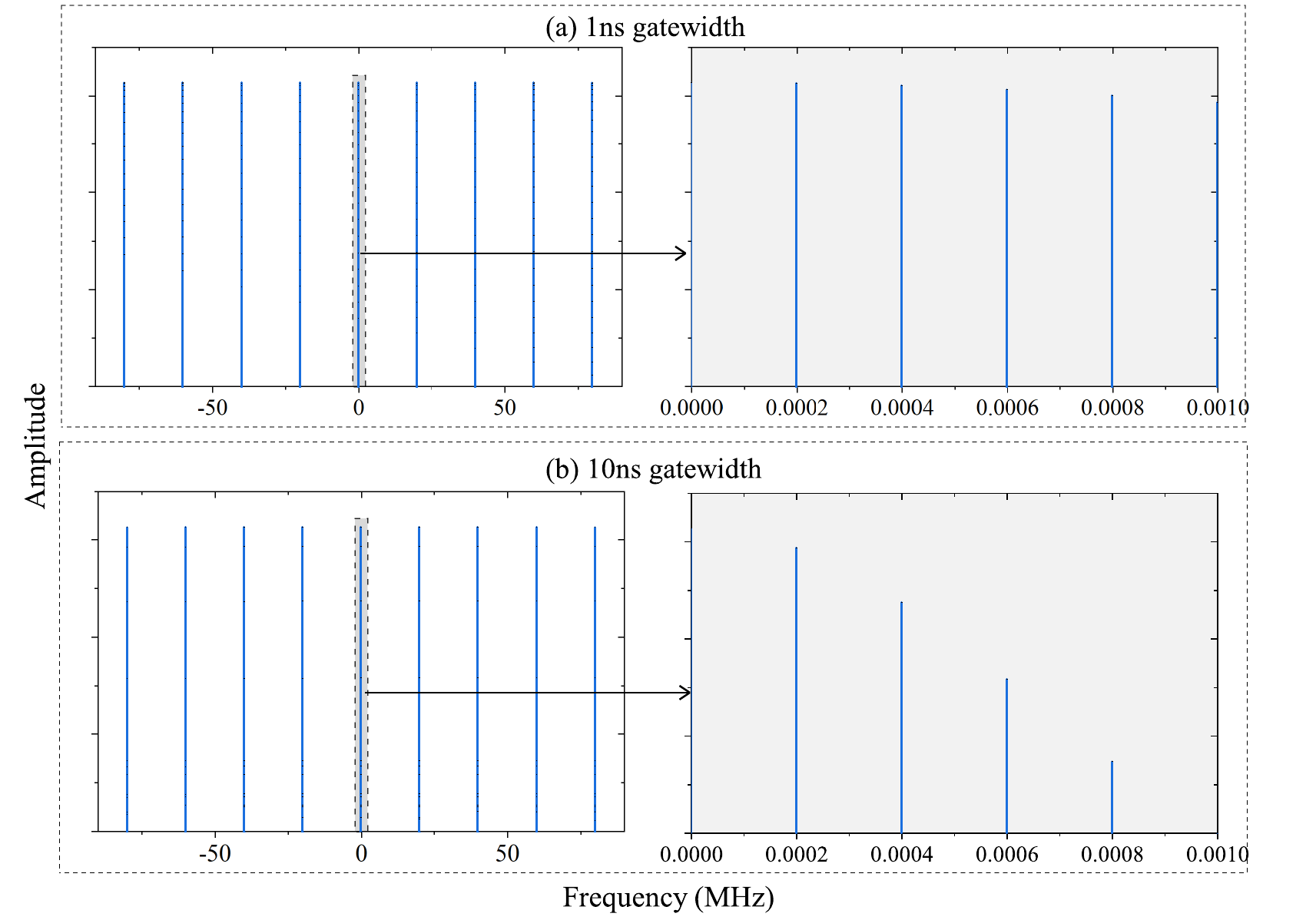 }
    \caption{\label{fig: FFT_gated_mode} \textbf{ Expected frequency spectrum of a gated-mode system with misaligned clock} Regarding Bob's clock as the reference 20 MHz, Bob opens his gate at 20 MHz and Alice prepares and sends the quantum state at 20.0002 MHz. The gray subfigures show a magnified view of the fine-structure components around the corresponding zero-frequency region. When gated-mode is employed, Bob can track Alice's clock by the fine-structure components. Subfigures (a) and (b) show that Bob's gate width are 1 ns and 10 ns respectively. Clearly, a larger gate width leads to faster attenuation of the fine-structure components. When gate width equals the system period, only a single spectral line remains here, which reduces to the free-running case.} 
\end{figure}

\subsection{ Gated-mode filtering and low-speed sampling}

Noting that the beat frequency $f_{bt} = |\Delta_f|$ is sufficient for the frequency recovery, the previous results enlighten us that we can exploit the beat phenomenon and employ our low-speed sampling strategy. Compared with the previous high-speed sampling, counting detection events over a time interval ${\tau_s'}$ exhibits the following characteristics. First, regarding the jitter $\chi_k$, $k\tau_A$ and $k\tau_A + \chi_k$ largely remain within the same ${\tau_s'}$ interval (only counts near the very edge of the interval can fall into the adjacent interval, and the  probability is negligible). Second, since the afterpulse probability decays exponentially, a SPD count at $k\tau_A$ and its afterpulse at $k\tau_A + T_d + \epsilon_k$ are most likely to occur within the same interval (neglecting higher-order afterpulses and considering the dead time, at most one afterpulse in a given interval can be attributed to a detection event from an adjacent interval). Finally, due to the same reason, the influence of dead time on the adjacent interval is also negligible. Based on these properties, we conclude that the proposed low-speed sampling method effectively mitigates the impact of dead time, afterpulse, and timing jitter. Therefore, the count number $x(l)$ for each time interval $l$ can be treated as approximately independent.

Supposing Bob has divided the total sampling time $T_s$ into $2M$ intervals, each of duration ${\tau_s'} = T_s / 2M$, and $L$ intervals form one beat period. For simplicity, assume that $L$ is an integer.
Due to the frequency mismatch, we assume that within one beat period $\tau_{bt} = L \tau_s' = 1/f_{bt}$, qubits fall within the detection window only during a small time interval $\tau_q$, while during the remaining time $\tau_v = \tau_{bt} - \tau_q$, all signals lie outside the detection window. Without loss of generality, we further assume that $\tau_q < {\tau_s'}$ and Alice’s pulses are perfectly aligned with Bob’s detection window at the initial time. Thus, within each beat period, an interval that received qubits has a higher expected count $n_q$, while the remaining $L-1$ intervals only have dark count level $n_v$. The sampling obtains a count number array ${\bm x}$ consists of $2M$ count numbers $x(l)$ where $l = {-M, -(M-1), \ldots, -1, 0, 1, \ldots, M-1}.$ The expected value of $x(l)$ is:
\begin{equation}
\langle x(l) \rangle =
\begin{cases}
{n_q}, & \text{if } l \bmod L = 0, \\
{n_v}, & \text{otherwise}.
\end{cases}
\end{equation}

Performing a Fourier analysis on $f(n)$ yields:
\begin{equation}
\label{eq. Xk1}
\begin{aligned}
&\langle X (k) \rangle = \sum_{l=-M}^{M-1} \langle x(l) \rangle e^{i2\pi k l / 2M} \\
&= \sum_{l=-M}^{M-1} n_v e^{i2\pi k n / 2M} + \sum_{r=-M/L}^{M/L - 1} (n_q - n_v)  e^{i2\pi k r L / 2M}.
\end{aligned}
\end{equation}
Given that $M \gg L$ and $M$ is large, the first term in Eq. (\ref{eq. Xk1}) satisfies 
\begin{equation}
\sum_{n=-M}^{M-1} n_v e^{i2\pi k l / 2M} =
\begin{cases}
2M n_v, & \text{if } k = 0, \\
0, & \text{otherwise}.
\end{cases}
\end{equation}

For the second term
\begin{equation}
    \begin{aligned}
        &\sum_{r=-M/L}^{M/L - 1} (n_q - n_v) e^{i2\pi k r L / 2M} \\
          =&\begin{cases}
            2M (n_q - n_v) /L,  \text{if } k \mod{(2M/L)} = 0, \\
            0,  \text{otherwise}.
            \end{cases}
    \end{aligned}
\end{equation}

Therefore,
\begin{equation}
\begin{aligned}
\langle {X}(k)  \rangle
=\begin{cases}
2M \left( {n_q}/{L} + {n_v(L-1)}/{L} \right), ~ \text{if } k = 0, \\
{2M(n_q - n_v)}/{L} , ~ \text{if } k \mod 2M/L = 0,\ k \neq 0, \\
0, ~ \text{otherwise},
\end{cases} 
\end{aligned}
\end{equation}
The peak position $2M/L$ indicates that the frequency misalignment is approximate ${2M}/{(T_s L)}$.

\subsection{Signal-to-noise ratio analysis}

When discussing SNR, we focus more on the power spectral density (PSD) since the phase is not important. As we have discussed, the count numbers $x(l)$ are approximately mutually independent.
We start from the asymptotic case, i.e., $M \rightarrow \infty $, the PSD can be expressed as 
\begin{equation}
\begin{aligned}
\label{Eq Var1}
& \langle S(k) \rangle = \langle |X(k)|^2 \rangle / 2M \\
& = \frac{1}{2M} \sum_{l=-M}^{M-1}\sum_{m=-M}^{M-1} \langle x(l)x(m) \rangle e^{i2\pi k(l-m)/2M}.
\end{aligned}
\end{equation}

As $x(l)$ for each time interval $l$ can be treated as approximately independent, it follows that
\begin{equation}
\langle x(l)x(m) \rangle =
\begin{cases}
\langle x(l)\rangle^2 + \sigma^2_{x_l} & \text{if } l = m, \\
\langle x(l)\rangle \langle x(m)\rangle & \text{if } l \neq m,
\end{cases}
\end{equation}
where $\sigma^2_{x_l} := \langle x(l)^2 \rangle - \langle x(l) \rangle^2  $ is the variance of $x(l)$. Thus, we obtain
\begin{equation}
    \begin{aligned}
        &\sum_{m = -M}^{M-1} \langle x(m)x(m+\tau) \rangle  \\
        &= \frac{4M}{L}  n_q   n_v  + \frac{2M(L-2)}{L}  n_v^2 \\
        &= 2M \Gamma_1, \quad \tau \in \mathbb{Z},~ \tau \neq qL,~ q \in \mathbb{Z},
    \end{aligned}
\end{equation}
where $\Gamma_1 := \frac{2}{L}  n_q   n_v  + \frac{(L-2)}{L}  n_v^2$,
and
\begin{equation}
        \begin{aligned}
            &\sum_{m = -M}^{M-1} \langle x(m)x(m+qL) \rangle \\
           &  = \frac{2M}{L}  n_q  ^2 + \frac{2M(L-1)}{L}  n_v^2 \\
            &= 2M\Gamma_2, \quad q \in \mathbb{Z},~ q \neq 0,
    \end{aligned}
\end{equation}
where $\Gamma_2 := \frac{1}{L}  n_q  ^2 + \frac{(L-1)}{L}  n_v^2$,
and
\begin{equation}
\sum_{m = -M}^{M-1} \langle x(m)^2 \rangle = 2M\Gamma_2 + \frac{2M}{L} \sigma_{q}^2 + \frac{2M(L-1)}{L} \sigma_{v}^2.
\end{equation}
where $\sigma_s^2$ and $\sigma_v^2$  are defined as $\sigma_{x_l}^2$ for $l \mod L = 0$ and $l \mod L \neq 0$ respectively.

Therefore, Eq.~\eqref{Eq Var1} can be rewritten as
\begin{equation}
\begin{aligned}
\label{Eq Var2}
 & \langle S(k) \rangle= \sum_{q = -M/L}^{M/L-1} (\Gamma_2 - \Gamma_1) e^{i2\pi k qL/2M} + \frac{1}{L} \sigma_{q}^2 \\
&+ \frac{L-1}{L} \sigma_{v}^2 + \sum_{\tau = -M}^{M-1} \Gamma_1 e^{i2\pi k \tau/2M},
\end{aligned}
\end{equation}
which further yields
\begin{equation}
\begin{aligned}
   & \langle S(0) \rangle = \frac{2M}{L}(\Gamma_2 - \Gamma_1)  
          + \frac{1}{L} \sigma_{q}^2 \\
         &\quad + \frac{L-1}{L} \sigma_{v}^2 
          + 2M \Gamma_1 =: S_{DC},
\end{aligned}
\end{equation}
for the direct component $k=0$;
\begin{equation}
\begin{aligned}
   & \langle S(k) \rangle = \frac{2M}{L}(\Gamma_2 - \Gamma_1)  
          + \frac{1}{L} \sigma_{q}^2 \\
         &\quad + \frac{L-1}{L} \sigma_{v}^2 =: S_{bf}, 
\end{aligned}
\end{equation}
for the beat frequency component and its harmonics $\quad k = qL$, $k \neq 0$, and $q \in \mathbb{Z}$; and 
\begin{equation}
\begin{aligned}
   \langle S(k)  \rangle = \frac{1}{L} \sigma_{q}^2 
          + \frac{L-1}{L} \sigma_{v}^2 = S_{noi}, \quad \text{otherwise}.
\end{aligned}
\end{equation}
for noisy frequency components.



\hfill

The SNR of the PSD is given by

\begin{equation}
\begin{aligned}
\label{Eq Var3}
&\frac{S_{bf}}{S_{noi}} =
\frac{2M(\Gamma_2 - \Gamma_1) + \sigma_{q}^2 + (L-1) \sigma_{v}^2 }
{\sigma_{q}^2 + (L-1) \sigma_{v}^2} \\
&= \frac{2M}{L}\frac{\left(n_{q} - n_{v}\right)^2}
{\sigma_{q}^2 + (L-1) \sigma_{v}^2}+1.
\end{aligned}
\end{equation}
We define $N'_{s} = \tau'_{s} / \tau_B $ as the QKD round number in one time interval, $N_{q} = \tau_q / \tau_B$ as the round number given that the qubit falls in the detection window in one time interval, and $N'_v = N'_{s} - N_{q}$. Meanwhile, we define $Q_q$ and $Q_v$ as the count rate given that the qubit falls in and falls out of the detection window respectively.
Then the expectations and variances satisfy
\begin{align}
n_{q} &\approx N_q Q_q + N'_v Q_v, \\
\sigma_{q}^2 &\approx N_q Q_q(1-Q_q) + N'_vQ_v(1-Q_v), \\
n_{v} &\approx N_s Q_v, \\
\sigma_{v}^2 &\approx N_s Q_v(1-Q_v),
\end{align}

Furthermore, noting that $Q_v \ll Q_s \ll 1$, Eq.~\eqref{Eq Var3} can be rewritten as
\begin{equation}
\begin{aligned}
\label{Eq Var4}
&\frac{S_s}{S_n} = \frac{2M}{L}
\frac{(N_qQ_q - N_qQ_v)^2}
{N_qQ_q(1-Q_q) + (LN'_s - N_q) Q_v(1-Q_v)}+1 \\
&\approx \frac{2M}{L}
\frac{(N_qQ_q - N_qQ_v)^2}
{N_qQ_q + (N_{bt} - N_q)Q_v}+1;
\end{aligned}
\end{equation}
where $N_{bt} = \tau_{bt}/\tau_B = LN'_s$.

It is worth noting that $ N_qQ_q - N_qQ_v+ LNQ_v =:n_{bt}$ exactly represents the expected count number within one beat period, and $n'_q = N_q (Q_q - Q_v) $ approximates the count number from qubits. The SNR can be further simplified as
\begin{equation}
\begin{aligned}
\label{Eq Var5}
\frac{S_{bf}}{S_{noi}} &= K \frac{{n_q'}^2}{n_{bt}}+1
= K r_q n'_q + 1,
\end{aligned}
\end{equation}
where $K = 2M/L$ denotes the total number of beat periods in the sampled data and $r_q = n_q'/n_{bt}$ represents the proportion of counts caused by qubits within one beat period.

\begin{figure}[htbp]
    \includegraphics[width=8cm]{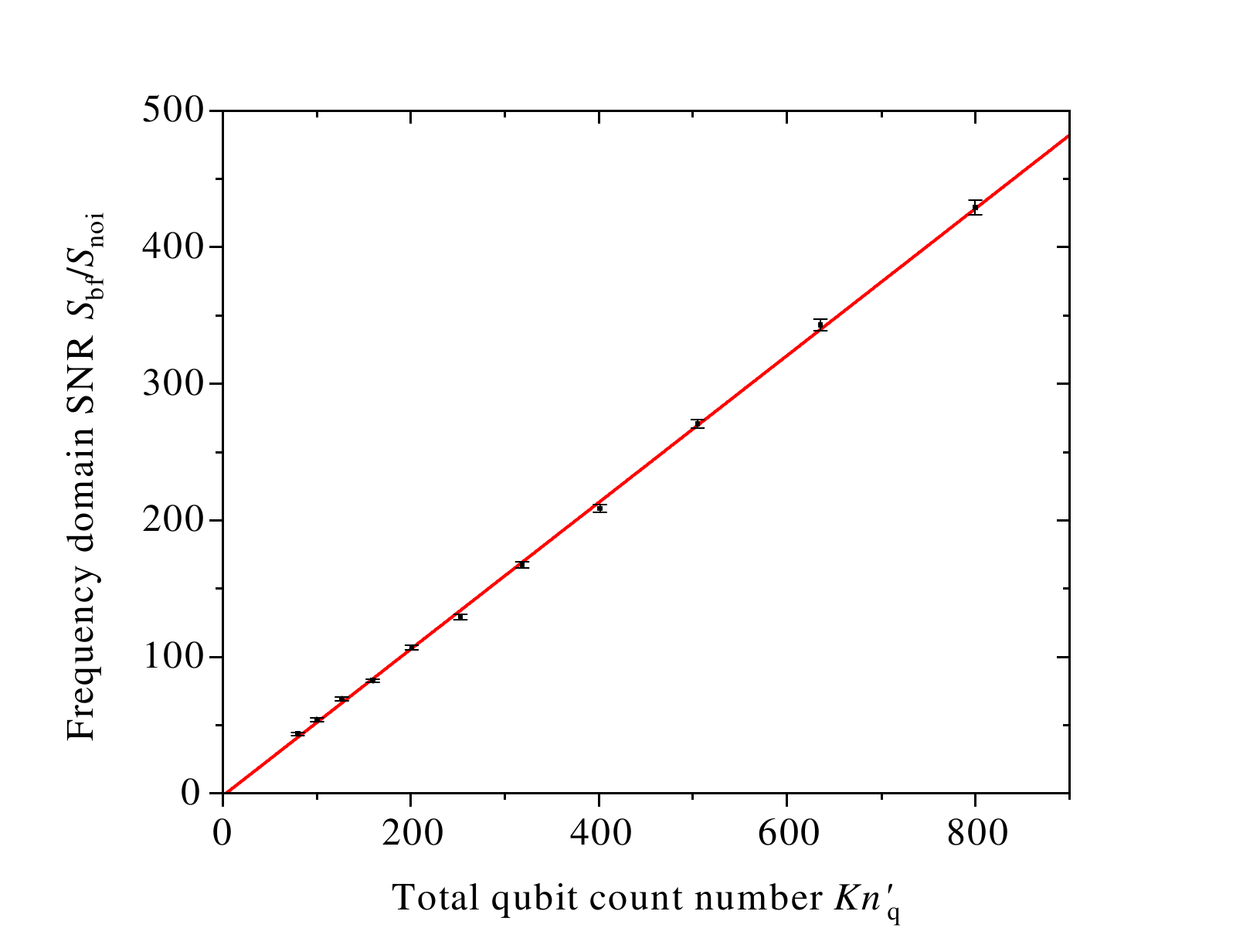 }
    \caption{\label{fig: SNR_VS_Nq} \textbf{Linear relationship between SNR and total qubit count number $Kn'_q$.} Here we fixed the sample time to 1 s ($K = 200$). $Kn'_q$ is increased from 80 to 800 with an increment of 1 dB per step while total noise count number $ K(n_{bt} - n'_q)$ is proportionally increased to maintain ${r} = 4/5$. $Kn'_q$ is increased from 80 to 800 with an increment of 1 dB per step. For each value of $Kn'_q$, we performed 50 Monte Carlo simulations and calculated the SNR for each run. The black squares in the figure represent the mean values and the standard error of the mean (SEM) of these 50 simulations, while the red line indicates the result of linear fitting.}
     \includegraphics[width=8cm]{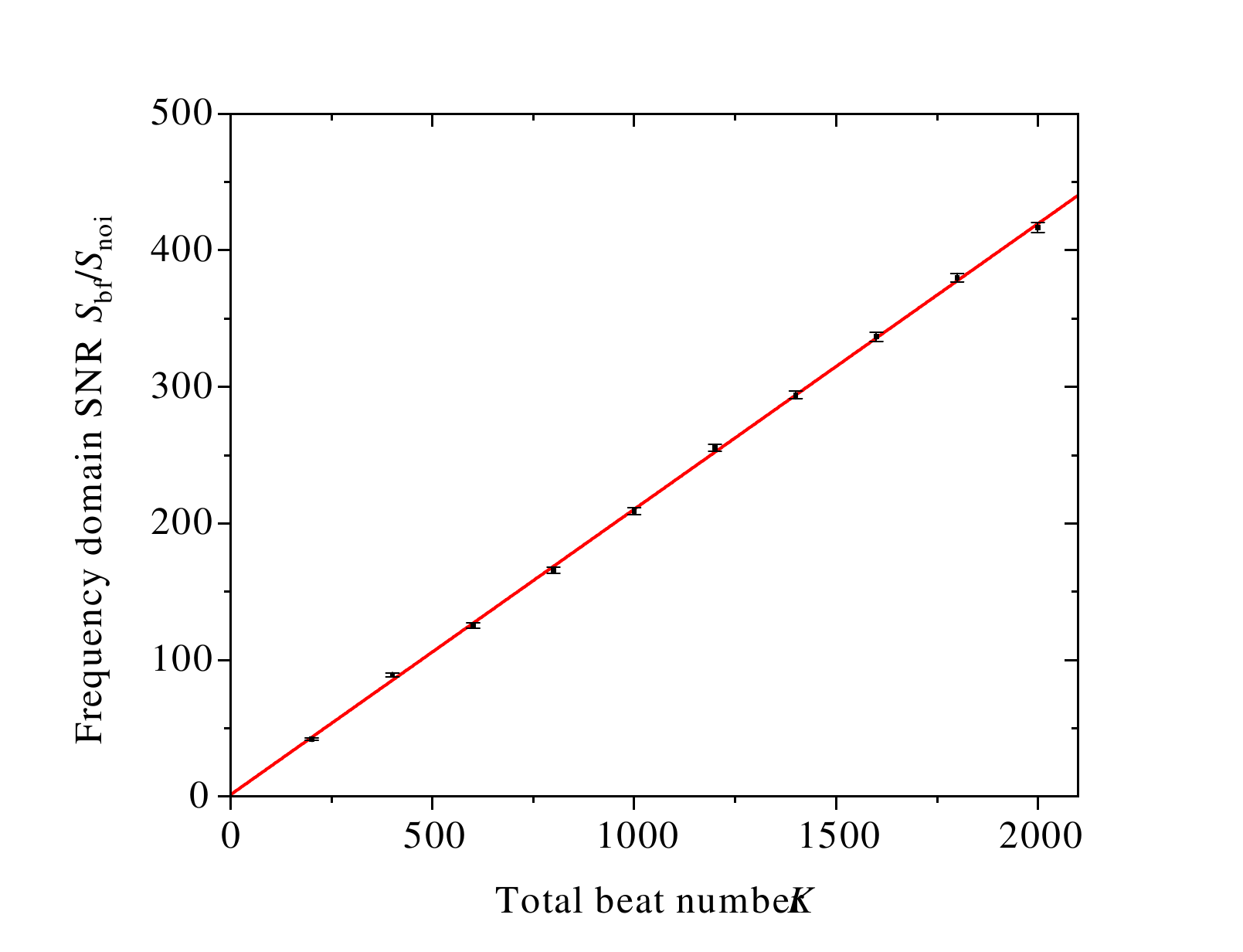 }
    \caption{\label{fig: SNR_VS_K} \textbf{Linear relationship between SNR and $K$.} Here we fixed expected $Kn'_q = 80$ and $\myExpect{N_{noi}} = 20$ and linearly increased sample time from 1 s ($K$ = 200) to 10 s ($K = 2000$). For each value of $K$, we performed 50 Monte Carlo simulations and calculated the SNR for each run. The black squares in the figure represent the mean values and the SEM of these 50 simulations, while the red line indicates the result of linear fitting.}
\end{figure}

As shown in Fig. \ref{fig: SNR_VS_Nq} and Fig. \ref{fig: SNR_VS_K}, we adopted a set of typical QKD parameters (20 MHz system frequency, 1 ns gate width, 200 Hz frequency mismatch, 1 $\rm {\mu s}$ dead time, 50 ps (std) jitter, and 2\% afterpulse rate) and used a controlled variable approach along with the Monte Carlo simulations to investigate the linear property of the SNR. The simulation results were then compared with theoretical predictions. 

\begin{figure}[htbp]
    \includegraphics[width=8cm]{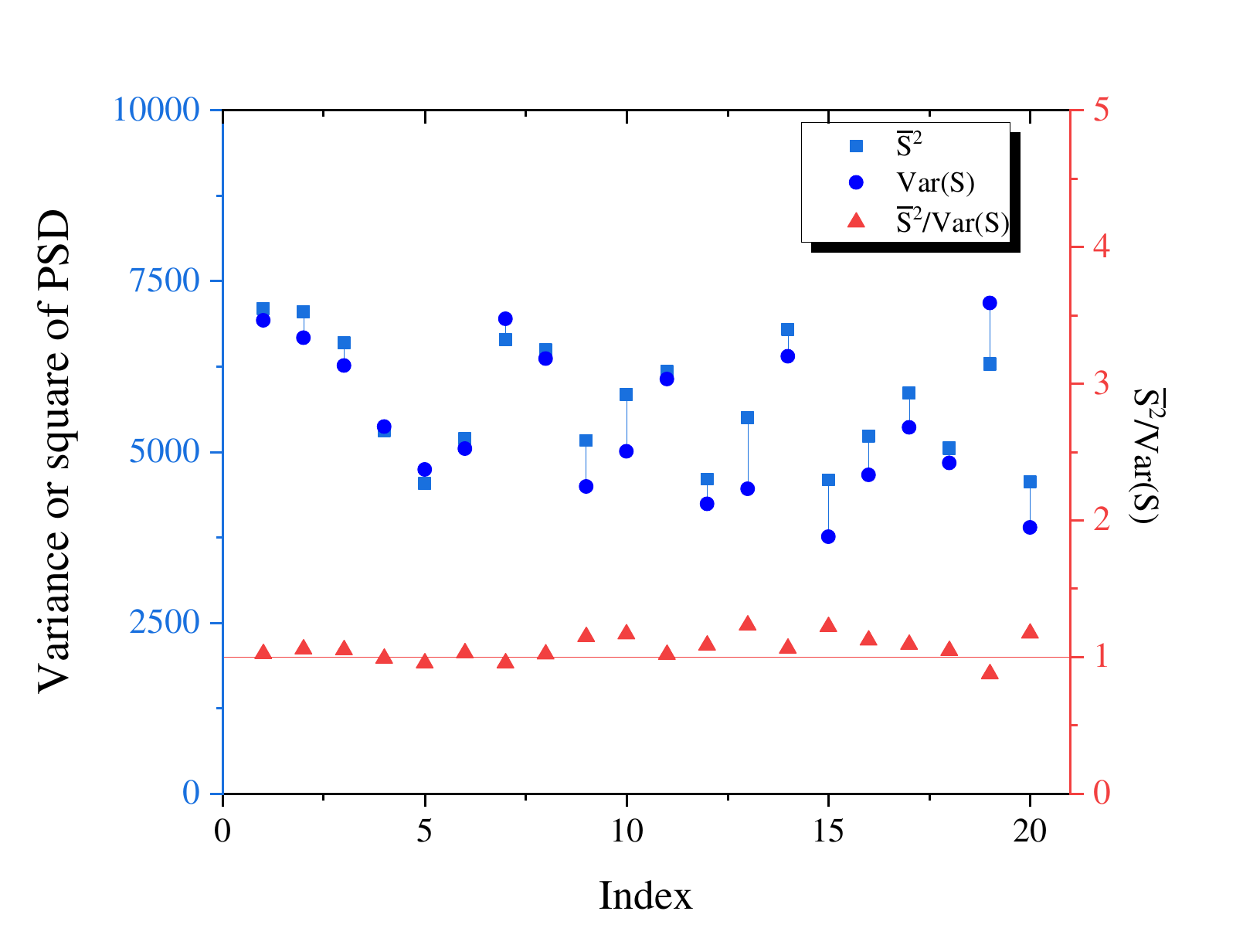 }
    \caption{\label{fig: variance of PSD}  \textbf{Variance and square of PSD results from Monte Carlo simulations.}  The blue squares and dots represent $\bar{S}_n^2$ and ${\text{Var}(S_n)}$ (for noisy frequency) respectively.
    The red triangles represent the ratio $\text{Var}(S_n) \big/ \bar{S}_n^2$.  }
\end{figure}

\begin{figure}[htbp]
    \includegraphics[width=8cm]{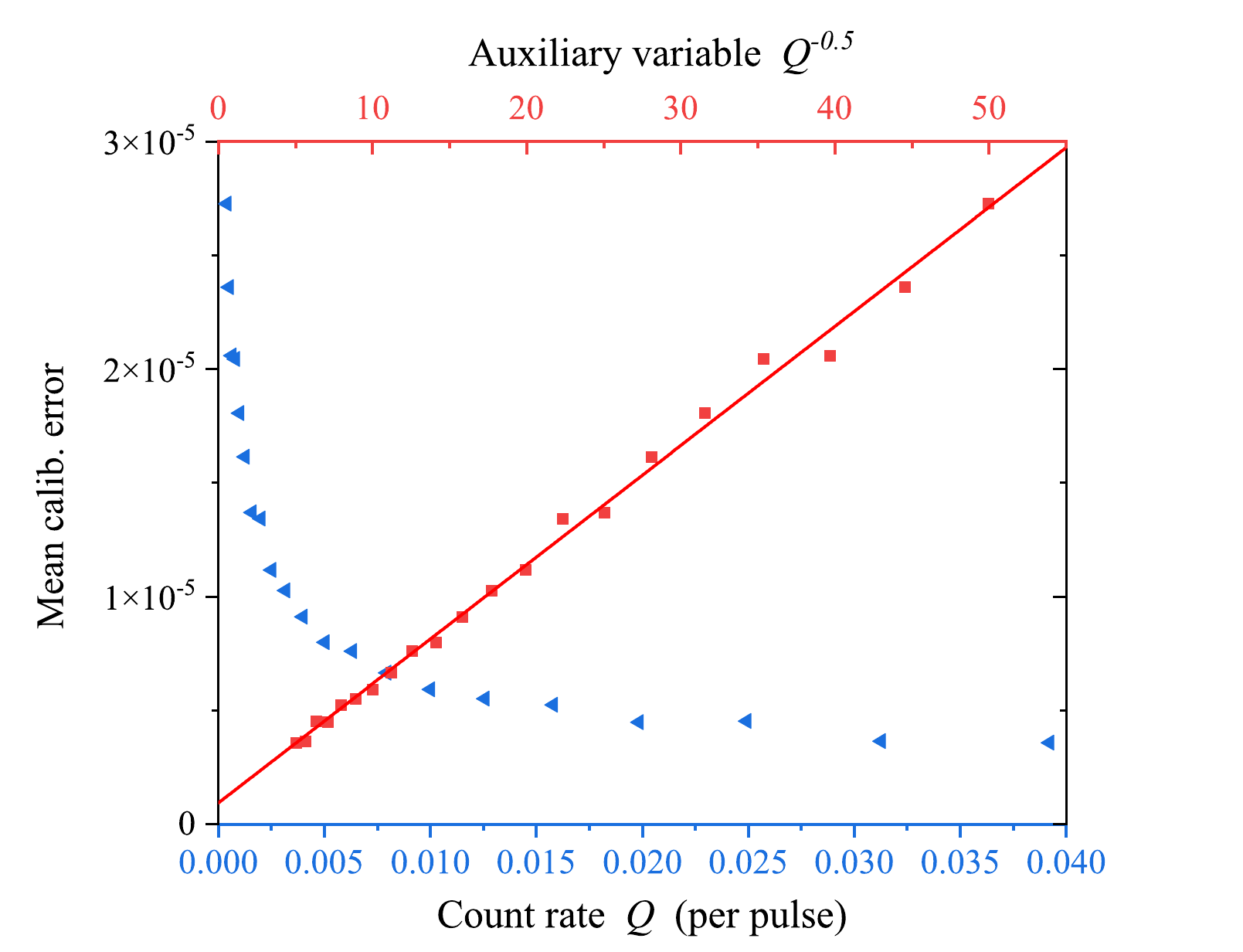 }
        \caption{\label{fig: initial calib}  \textbf{ Monte Carlo simulation of the initial frequency calibration} The Y-axis means the mean value of the calibration error $\sigma_{fin}$. The blue and red axis represent the qubit count rate $Q$ and the temporarily defined auxiliary variable $Q^{-\frac{1}{2}}$. The blue and red dots represent the mean $\sigma_{fin}$ versus the $Q$ and $Q^{-\frac{1}{2}}$ respectively. The red line represent linear fits to the corresponding data points of $Q^{-\frac{1}{2}}$, which indicates the linear relation between $Q^{-\frac{1}{2}}$ and $\sigma_{fin}$. }
\end{figure}

Regarding the variance of the PSD at noisy frequencies, when the time-domain samples follows a Gaussian distribution, Isserlis' theorem \cite{isserlis1918formula,vignat2012generalized} ensures 
\begin{equation}
    \sigma_{S}^2(f) = \langle {S(f)} \rangle ^2,\ \ \text{for $f\neq 0$}
\end{equation}
which means that the variance of the PSD $\sigma_{S}^2(f)$ at any non-zero frequency $f$ equals the square of its mean PSD value $\langle S(f) \rangle ^2$. In non-Gaussian cases, higher-order moment corrections are required, so it is difficult to provide an elegant closed form with current theoretical tools struggle. However, the central limit theorems guarantee that with a sufficiently large number of rounds, the photon count sampling in each time interval will approximate a Gaussian distribution. Therefore, we approximately equate the standard deviation of the PSD 
$\sigma_{S}(f)$ to $\langle {S(f)} \rangle$ in our scenario. We ran the Monte Carlo simulation 20 times and obtained 20 sets of PSD data to calculate the average square value $\bar{S}_n^2$ and variance ${\text{Var}(S_n)}$ of the SPD value at the noisy frequencies, where the subscript $n$ indicates the $n$-th simulation. Here, our $\bar{S}_n$ and $\text{Var}(S_n)$ represent the mean and variance of $S_n(f)$ over the noisy frequency $f$, rather than the individual mean and variance of each $S(f)$. Although it is mathematically challenging to rigorously prove the equivalence of this approach, it nevertheless ensures sufficient accuracy for practical engineering applications. The results are shown in Fig. \ref{fig: variance of PSD}. From the 20 sets of simulation we obtain $\text{Var}(S_n) \big/ \bar{S}_n^2 = 1.0661 \pm 0.0913$, providing approximate support for our conclusion.

\subsection{Resolution analysis and frequency fine-tuning}

Practical Fourier analysis relies on the FFT, whose resolution $\Delta_{f_{coa}} = 1/T_s$ depends on the sample duration $T_s$, and the period resolution is $\Delta_{\tau_{coa}} = \tau_s / N_s$, where $N_s$ is the time interval number in the sample duration. If the sample duration $T_s$ is limited, the FFT result $\hat{\tau}_{coa}$ should be regarded as a coarse tuning, and statistical algorithms should be employed as a fine-tuning. The final resolution mainly depends on the statistical result $\hat{\tau}_{fin}$, and the resolution with practical issues is analyzed as follows. 

Given that most of the dark counts and afterpulses have been suppressed via gated-mode operation or temporal filtering, we primarily focus on the impact of timing jitter. Supposing the time jitter $\chi_j \sim \mathit{N}(0,\sigma^2)$ and $\chi_j$ with different $j$ are independent and identically distributed.  Equation (\ref{eq LSR})  demonstrates that the variance of the fine-tuning $\hat{\tau}_{fin}$ satisfies
\begin{equation}
\label{eq var}
\begin{aligned}
  \langle \text{Var}(\hat{\tau}_{fin}) \rangle =  
  \langle  \frac{2\sigma^2}{ \sum_j(k'_j - \bar{k}')^2} \rangle =
\frac{24\sigma^2}{Q N_s^2}
\end{aligned}
\end{equation}
where the numerator $2\sigma^2$ denotes the variance of $\text{Var}(\varepsilon'_j)$. Taking the standard deviation as the resolution of the fine-tuning, we get
\begin{equation}
\label{eq std}
\begin{aligned}
  \langle \sigma_{fin} \rangle \propto  \frac{\sigma}{Q^{0.5}{N_s}^{1.5}}   \ll  \Delta_{\tau_{coa}},
\end{aligned}
\end{equation}
which indicates a much higher accuracy for the robust clock synchronization.

To validate Eq. (\ref{eq std}), we use the aforementioned Monte Carlo simulation to simulate our Fourier analysis with different initial clock misalignments at different qubit count rate. For each qubit count rate, we randomly generate 500 different frequency misalignments following the uniform distribution in $[-10^4\text{ Hz},10^4\text{ Hz}]$. For each frequency misalignment, we simulate 1 s ($2\times10^7$ rounds) of QKD operational data, execute the initial Qubit4Sync (Modu.1 $\rightarrow$ Modu.3 $\rightarrow$ Modu.4 $\rightarrow$ Modu.5) with this data, and record the absolute calibration error (i.e., the absolute difference between the estimated and true frequency offsets). The simulation results are shown in Fig. \ref{fig: initial calib}, where each blue square represents the mean absolute calibration error across 500 simulations at the corresponding fiber length. We further simulate the SPD yield $Q$ at each distance and use red triangles to represent the relationship between the mean absolute calibration error and $Q^{-0.5}$. The red fitting curve indicates the linear relationship and validates Eq. (\ref{eq std}). It can help Bob to determine how much data is required to ensure sufficient precision.

\bibliography{citations}

\end{document}